\documentclass{article}

\usepackage{arxiv}

\usepackage[utf8]{inputenc} 
\usepackage[T1]{fontenc}    
\usepackage[colorlinks=true,urlcolor = black,citecolor=black,linkcolor=blue]{hyperref}
\usepackage{hyperref}
\usepackage{url}            
\usepackage{booktabs}       
\usepackage{amsfonts}       
\usepackage{nicefrac}       
\usepackage{microtype}      

\usepackage{lipsum}         
\usepackage{graphicx}
\usepackage{doi}

\usepackage[utf8]{inputenc}
\usepackage{amsmath}
\usepackage{amssymb}
\usepackage{color}
\usepackage{amssymb,amsfonts,amsmath,amsthm,amsbsy} 
\usepackage{tabularx} 
\usepackage{mathtools,mathrsfs}
\usepackage{mhchem}
\usepackage{enumerate}
\usepackage{booktabs}
\usepackage{bm}
\usepackage{bbm}
\usepackage{multirow}
\usepackage{threeparttable}
\usepackage{float}

\usepackage[table,dvipsnames, svgnames]{xcolor}

\usepackage[stable]{footmisc}
\usepackage[sort&compress]{natbib}

\usepackage{indentfirst}
\usepackage{etoolbox,siunitx}

\usepackage{afterpage}
\robustify\bfseries
\usepackage{adjustbox}
\usepackage{chngpage}
\usepackage{rotating}
\usepackage{hvfloat}
\usepackage{soul}
\usepackage{soul}
\RequirePackage{color}

\usepackage{subcaption}
\usepackage{caption}

\usepackage{cleveref}       
\title{Mechanism of Cold-spot Autoignition in a Hydrogen/Air Mixture}

\author{ \href{https://orcid.org/0000-0002-4788-9877}{\includegraphics[scale=0.06]{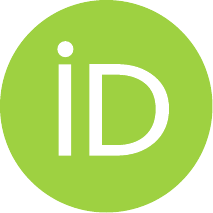}\hspace{1mm}Dimitris M. Manias\thanks{Corresponding Author: dimitris.manias@ku.ac.ae}}\\
	Department of Mechanical Engineering\\
	Khalifa University of Science and Technology\\
	Abu Dhabi, UAE 127788 \\
	\texttt{dimitris.manias@ku.ac.ae} \\
	\And
Wonsik Song \\
	Department of Energy and Process Engineering\\
	Norwegian University of Science and Technology\\
	Høgskoleringen 1, 7034 Trondheim, Norway\\
	\texttt{wonsik.song@ntnu.no} \\
	\And
	\href{https://orcid.org/0000-0003-2894-1067}{\includegraphics[scale=0.06]{orcid.pdf}\hspace{1mm}Aliou Sow} \\
	Department of Mechanical Engineering\\
	University of Ottawa\\
	161 Louis Pasteur, Ottawa, ON, Canada K1N6N5\\
	\texttt{asow2@uottawa.ca} \\
	\And
	\href{https://orcid.org/0000-0001-8849-1251}{\includegraphics[scale=0.06]{orcid.pdf}\hspace{1mm}Efstathios-Al. Tingas}\\
School of Computing, Engineering\\
 and the Built Environment\\
Edinburgh Napier University\\
	Edinburgh EH10 5DT, UK\\
	\texttt{e.tingas@napier.ac.uk} \\
	\And
	\href{https://orcid.org/0000-0001-7906-8646}{\includegraphics[scale=0.06]{orcid.pdf}\hspace{1mm}Francisco E. Hernandez Perez}\\
	Clean Combustion Research Center\\
	King Abdullah University of Science and Technology\\
	Thuwal 23955-6900, Saudi Arabia\\
	\texttt{francisco.hernandezperez.1@kaust.edu.sa} \\
	\And
	\href{https://orcid.org/0000-0001-7080-1266}{\includegraphics[scale=0.06]{orcid.pdf}\hspace{1mm}Hong G. Im}\\
	Clean Combustion Research Center\\
	King Abdullah University of Science and Technology\\
	Thuwal 23955-6900, Saudi Arabia\\
	\texttt{hong.im@kaust.edu.sa} \\
	\And
	\href{https://orcid.org/0000-0002-9674-9491}{\includegraphics[scale=0.06]{orcid.pdf}\hspace{1mm}Dimitris A. Goussis}\\
Research and Innovation Center on CO$_2$ and H$_2$\\
  Department of Mechanical Engineering\\
	Khalifa University of Science and Technology\\
 Abu Dhabi 127788, United Arab Emirates\\
 \texttt{dimitris.goussis@ku.ac.ae} \\
}

\renewcommand{\headeright}{Research Paper}
\renewcommand{\undertitle}{Completed Research Paper}
\renewcommand{\shorttitle}{\textit{Mechanism of Cold-spot Autoignition in a Hydrogen/Air Mixture} }

\hypersetup{
pdftitle={Mechanism of Cold-spot Autoignition in a Hydrogen/Air Mixture},
pdfsubject={physics.chem-ph},
pdfauthor={Dimitris M. Manias},
pdfkeywords={Detonation, Pre-ignition, Hydrogen, Computational Singular Perturbation, Cold-spot Autoignition},
}

\begin{document}
\maketitle

\begin{abstract}
When designing high-efficiency spark-ignition (SI) engines to operate at high compression ratios, one of the main issues that have to be addressed is detonation development from a pre-ignition front. In order to control this phenomenon, it is necessary to understand the mechanism by which the detonation is initiated. The development of a detonation from a pre-ignition front was analyzed by considering a one-dimensional constant-volume stoichiometric hydrogen/air reactor with detailed chemistry. A spatially linear initial temperature profile near the end-wall was employed, in order to account for the thermal stratification of the bulk mixture. A flame was initiated near the left wall and the effects of its propagation towards the cold end-wall were analyzed. Attention was given on the autoignition that is manifested within the cold-spot ahead of the flame and far from the end-wall, which is followed by detonation. Using CSP tools, the mechanism by which the generated pressure waves influence the autoignition within the cold-spot was investigated. It is found that the pressure oscillations induced by the reflected pressure waves and the pressure waves generated by the pre-ignition front tend to synchronize in the chamber, increasing the reactivity of the system in a periodic manner. The average of the oscillating temperature is greater in the cold-spot, compared to all other points ahead of the flame. As a result, the rate constants of the most important reactions are larger there, leading to a more reactive state that accelerates the dynamics of the cold-spot and to its autoignition.
\end{abstract}


\section{Introduction}
\label{Introduction}

When designing high-efficiency spark-ignition (SI) engines to operate at high compression ratios, one of the main issues that have to be addressed is detonation development from a pre-ignition front. In order to control this phenomenon, it is necessary to understand the mechanism by which the detonation is initiated. The development of a detonation from a pre-ignition front was analyzed by considering a one-dimensional constant-volume stoichiometric hydrogen/air reactor with detailed chemistry. A spatially linear initial temperature profile near the end-wall was employed, in order to account for the thermal stratification of the bulk mixture. A flame was initiated near the left wall and the effects of its propagation towards the cold end-wall were analyzed. Attention was given on the autoignition that is manifested within the cold-spot ahead of the flame and far from the end-wall, which is followed by detonation. The mechanism by which the generated pressure waves influence the autoignition within the cold-spot was investigated. 

It is found that the pressure oscillations induced by the reflected pressure waves and the pressure waves generated by the pre-ignition front tend to synchronize in the chamber, contributing to the preconditioning of the end-gas that becomes highly reactive, resulting in a local pressure build-up and acceleration of the flame due to a higher chemical heat release rate.~As a result, the reactivity of the system is increased in a periodic manner. The average of the oscillating temperature is greater in the cold-spot, compared to all other points ahead of the flame. Subsequently, the rate constants of the most important reactions become larger there, leading to a more reactive state that accelerates the dynamics of the cold-spot and to its autoignition.~Then, the chemical heat release of the autoignition is synchronized with the pressure pulses resulting in a detonation combustion mode. The detonation is characterized by a sharp increase in thermodynamic properties across the reactive wave, as well as a peak pressure that is much higher.

By using the CSP diagnostic tools, the species that relate the most to the explosive modes and the reactions that promote or oppose the generation of the explosive dynamics at the region of cold spot were identified and the relation with the oscillating pressure build-up was studied. In this study, the effect of transport was not taken under consideration, since this analysis concentrates on the effect of the chemistry involved on the evolution of the system.~A complete study, including the effect of transport to the evolution of the system will be conducted in a later study.

\section{Description of the Problem\footnote{Part of the text obtained from \citep{sow2019detonation}.}}
\subsection{Detonation Onset}
\label{Detonation}
Parametric studies have revealed that the temperature gradient in the bulk mixture clearly has an effect in the run-up time and intensity of the detonation. For different temperature gradients, from negative (cold spot) to positive (hot spot), the qualitative behavior of detonation development has been found to be different in the bulk mixture. For the case of a positive gradients, autoignition starts at the end-wall and triggers the detonation development due to the increased reactivity in the end gas. For the case of negative gradients, however, detonation development follows a typical pattern of DDT phenomena resulting from a self-acceleration of the deflagration front. The mechanism of the self reinforcement of the front is attributed to the increased residence time combined with the interactions between the incident and reflected pressure waves. As a secondary effect, the negative temperature slope creates a higher energy density in the bulk mixture, further contributing to the increased detonation intensity. Furthermore, for the negative temperature gradient case, an additional presence of a uniform temperature region just in front of the propagating flame results in yet another unique pattern of detonation development, in that autoignition starts within the colder region far from the wall, and later transitions to a detonation.

\subsection{Autoignition within the cold spot}

In the case of a negative temperature slope, the pressure pulses that are reflected from the wall and collide with the pressure pulses that are continuously generated by the propagating flame result in a the pressure and temperature build up locally where a local maximum for a very short period of time is exhibited. Immediately after, the waves move away, leading to a small drop in pressure and temperature. Meanwhile, pressure pulses generated by the propagating flame contribute to increase the pressure and temperature. When the reflected waves interact again with the pressure pulses, a local maximum is reinforced at a higher value than the previous one. These processes repeat at an accelerated pace due to the fact that the flame becomes more intense because of the higher end-gas reactivity, resulting in pressure pulses at higher strength. In addition, the distance between the flame and the end-wall is reduced as the process evolves. The pressure and the temperature become high and autoignition happens at 366.9 $\mu$s. From this observation, it is evident that the detonation development in the negative temperature slope conditions results from the repeated pressure wave interaction and reinforcement.

\section{\textbf{Computational Setup}}
\label{problem}

\begin{figure}[h]
\centering
\includegraphics[scale=0.9]{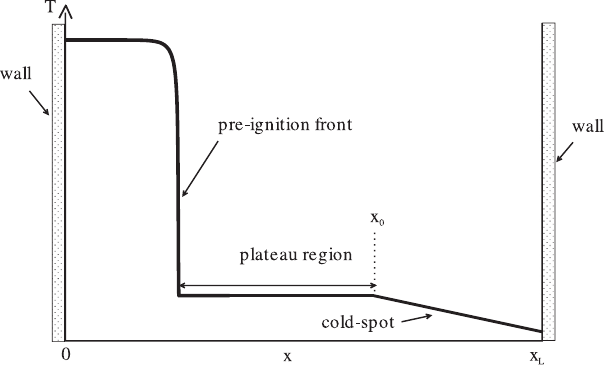}
\caption{Schematic representation of initial temperature distribution. Figure obtained from Aliou et al. \citep{sow2019detonation}.}
\label{fig:setup}
\end{figure}

The data calculated according to the configuration of Aliou et al.~\citep{sow2019detonation}.~The computational configuration of a one-dimensional constant volume reactor with impermeable and adiabatic walls at both boundaries was used.~A spatially linear initial temperature profile near the end-wall was employed, in order to account for the thermal stratification of the bulk mixture (see Fig.~\ref{fig:setup}).~The combustion chamber size, $x_L$, was set to 20 mm. For the initial condition, the Cantera \citep{goodwin2016cantera} solution for a freely-propagating premixed hydrogen–air flame with detailed transport properties was mapped near the left wall to represent the initial flame front developed by the pre-ignition event.~The flame travels from the left to the right where the initial pressure and velocity are constant and set to 10 atm and 0 m/s, respectively. Only stoichiometric conditions were considered in this study. The thermal stratification was modeled as a linear function. The magnitude of the temperature gradient, $dT/dx$, was taken 4 K/mm, which corresponds to the temperature difference of 16 K over the initial temperature, $T_0$ = 1030 K. The initial position, $x_0$, where the temperature variation starts, was taken at 16 mm.~A reactive stoichiometric hydrogen-air mixture, with a mechanism involving 9 reactive species and 23 elementary reaction steps, validated at high pressure conditions \citep{burke2012comprehensive}, was employed.

\section{Mechanism of the physical problem}
\label{Problem}

\begin{figure}[h]
\centering
\includegraphics[scale=0.5]{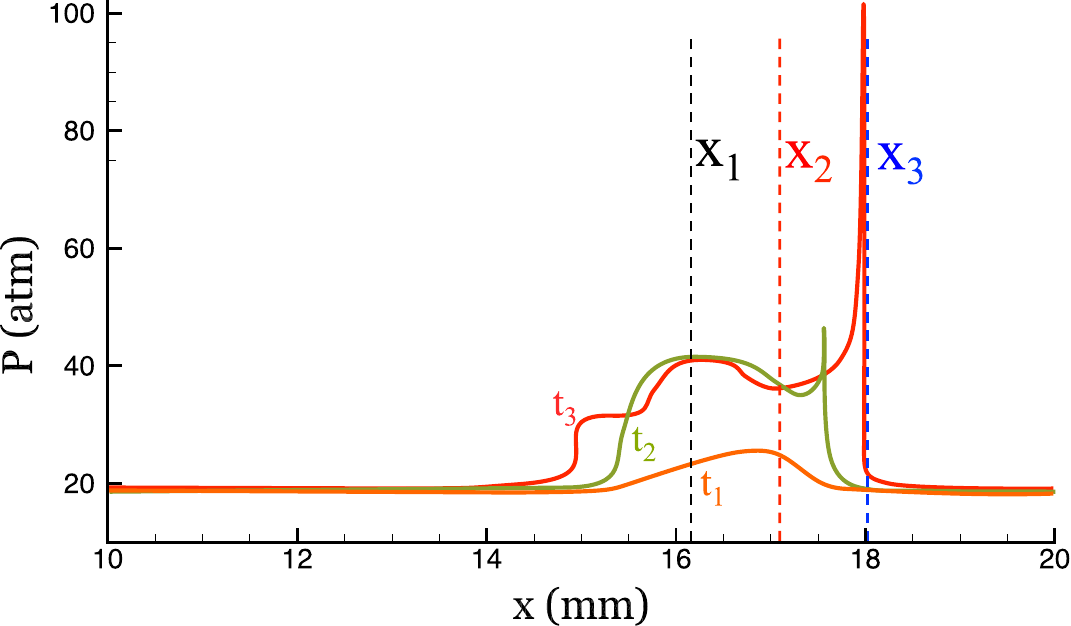}
\caption{Selection of the three points under investigation on the profile of pressure evolution.~Black line: point x$_1$, Red line: x$_2$, Blue line: x$_3$.}
\label{fig:points}
\end{figure}

In order to investigate the mechanism by which the ignition occurs inside the cold-spot, three indicative points were selected in the cold-spot region in the vicinity of the point where the autoignition occurs and their dynamics is analyzed by using tools derived from the CSP methodology.~In particular, as depicted of Fig.~\ref{fig:points}, the first point, x$_1$ = 16.2 mm, was selected in the region right after the start of the temperature stratification, the second point, x$_2$ = 17.1 mm, is the one where autoignition occurs and, finally, the third point, x$_3$ = 18 mm, was selected so that x$_2$ to be equidistant from x$_1$ and x$_3$, to the direction of the end-wall.~Note that x$_3$ is among the points subjected to the detonation initiation.\\

\subsection{Oscillatory behavior}
\subsubsection*{Pressure and Temperature Profiles}

\begin{figure}[h!]
\centering
\hfill
\begin{subfigure}[b]{0.45\textwidth} 
\includegraphics[scale=0.29]{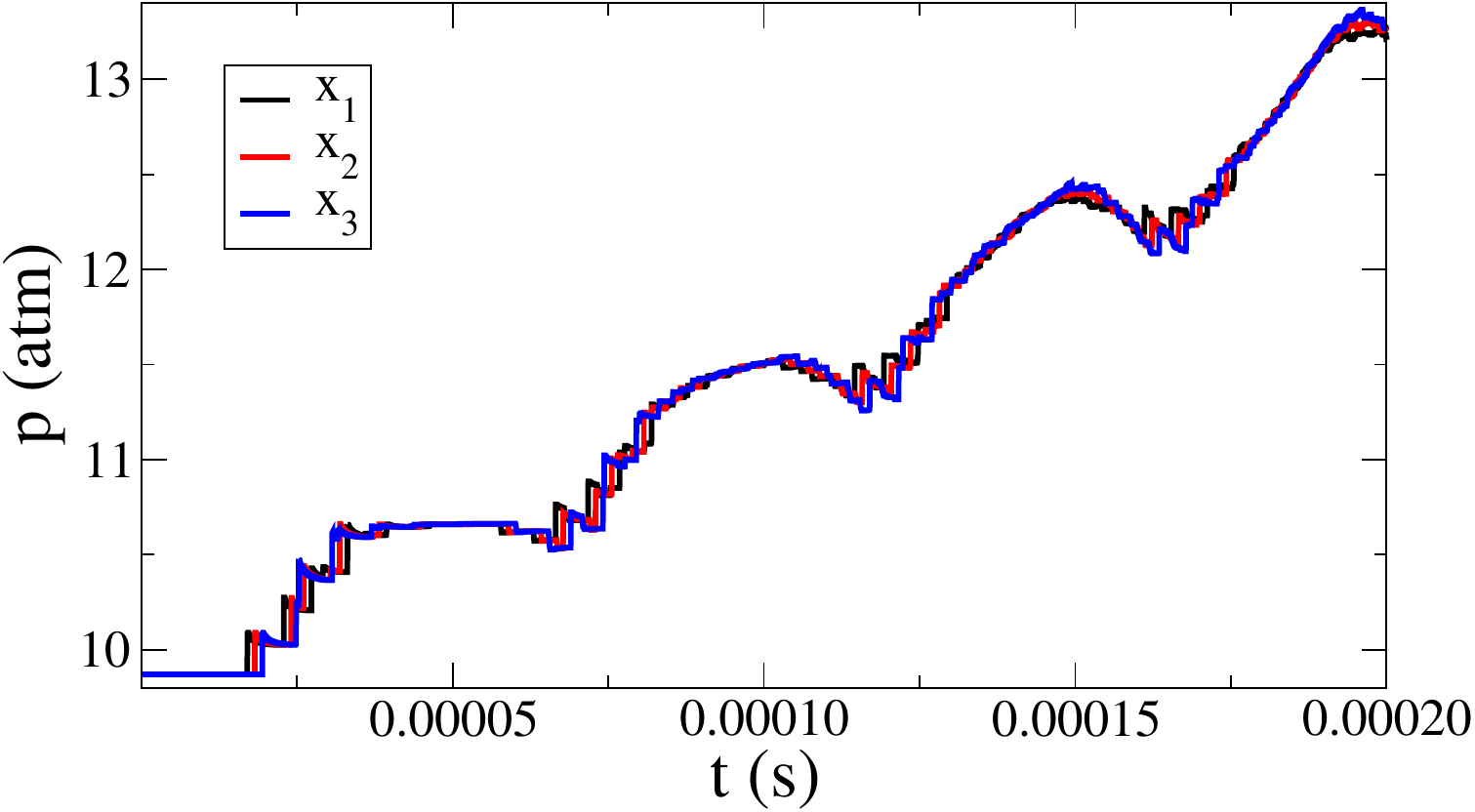}
\caption{}
\label{fig:P1}
\end{subfigure}
\hfill
\begin{subfigure}[b]{0.45\textwidth} 
\includegraphics[scale=0.29]{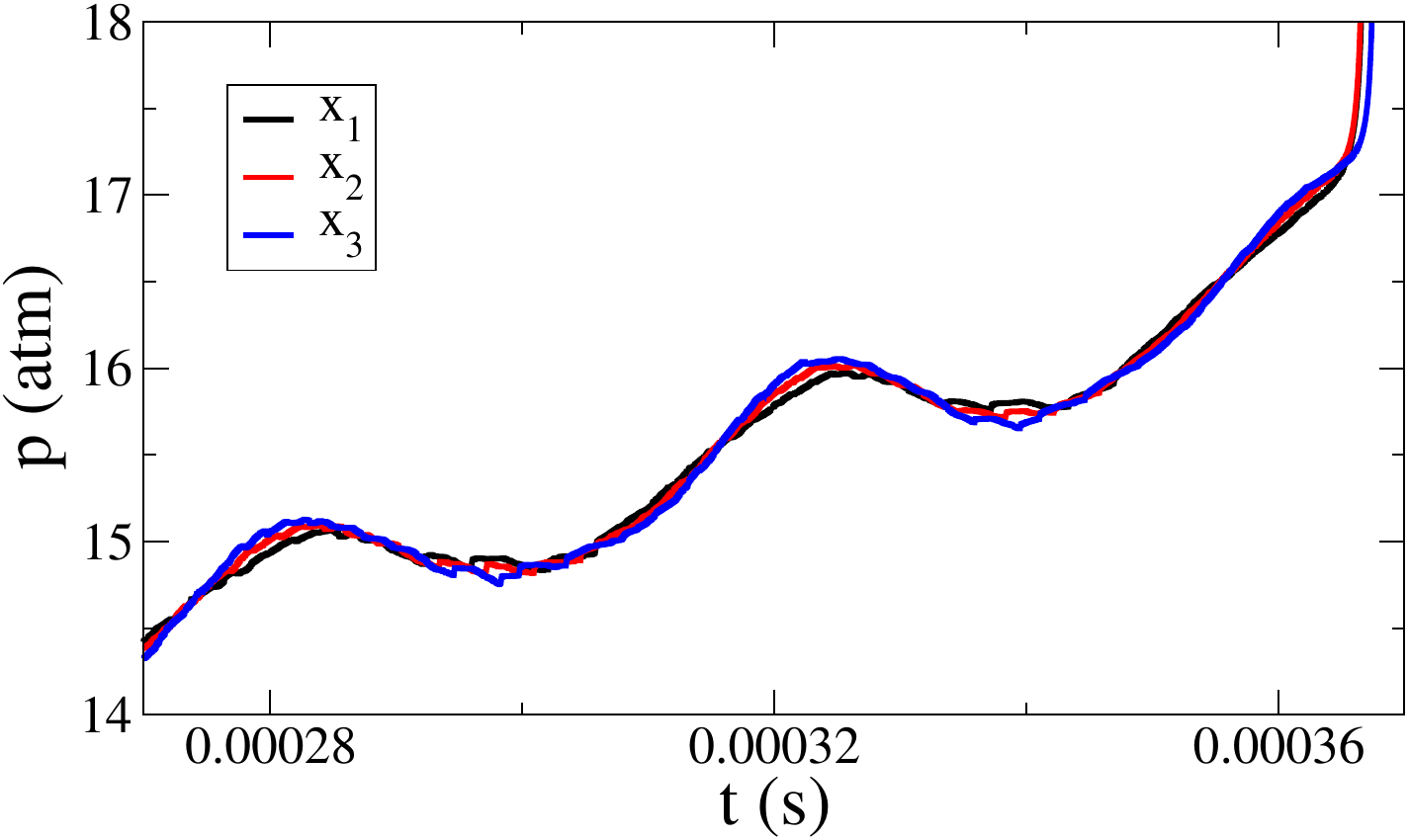}
\caption{}
\label{fig:P2}
\end{subfigure} 
\hfill ~
\\
\hfill
\begin{subfigure}[b]{0.45\textwidth} 
\includegraphics[scale=0.29]{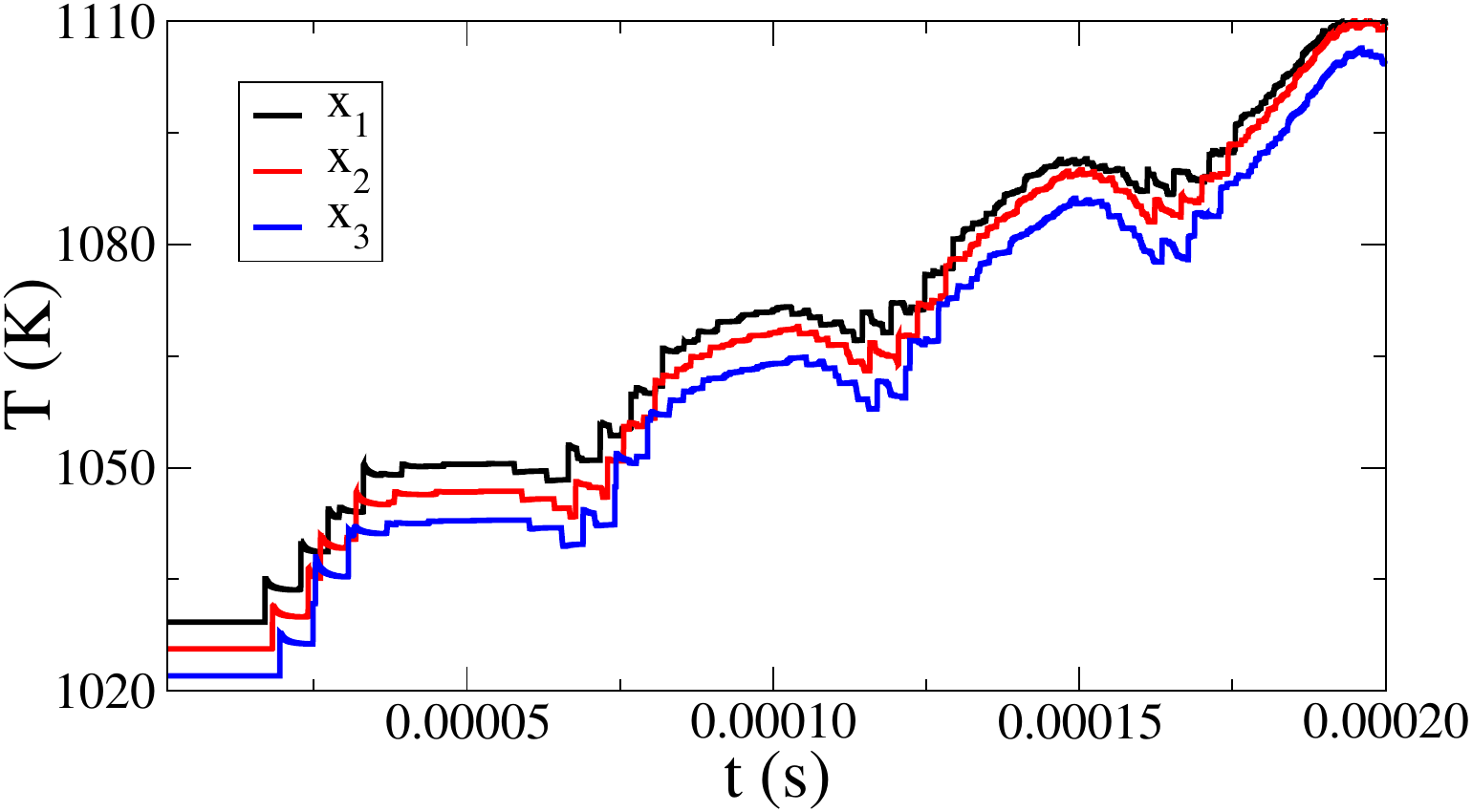} 
\caption{}
\label{fig:T1}
\end{subfigure}
\hfill
\begin{subfigure}[b]{0.45\textwidth} 
\includegraphics[scale=0.29]{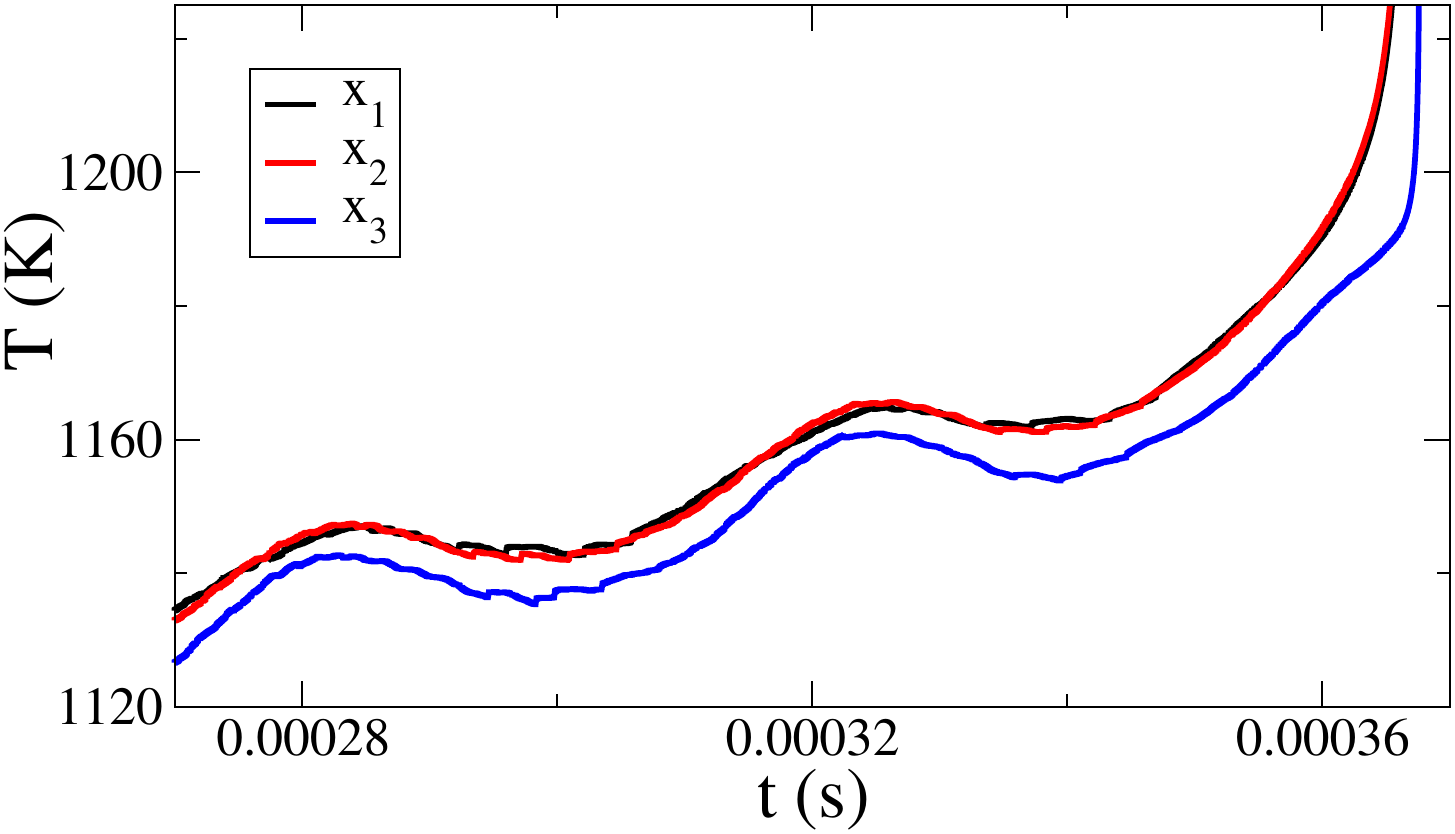}
\caption{}
\label{fig:T2}
\end{subfigure}
\hfill~
\caption{The pressure and temperature profiles for the three points under investigation.}
\label{fig:profiles}
\end{figure}

The pressure pulses generated by the pre-ignition front and the ones that have already been generated and reflected to the right boundary increase the pressure of the points of the end-gas region when reaching them. Initially, the effect of each pulse is clear; the points are affected one by one and try to revert to their initial state, as shown in Fig. \ref{fig:P1}, for the three points under investigation.~Until the next generated or reflected pulse approaches again, all points remain stationary at the last distorted state.~This repeated behavior continues until $\sim$250 $\mu$s, where the pressure pulses start to synchronize and every point in the domain follows eventually the same oscillatory pattern.~Figure \ref{fig:P2} suggests that, after 270 $\mu$s, the synchronized oscillations become more steep.~Additionally, it is evident that after $\sim$270 $\mu$s the magnitude of the oscillation differs significantly for every point.~Figure \ref{fig:P2} indicates that while moving towards the right wall, the magnitude of pressure oscillation becomes larger.\\

The same behavior of that of the pressure appears to the temperature profile, where the initial different forced increments of the temperature (Fig. \ref{fig:T1}) evolve into synchronized oscillations (Fig. \ref{fig:T2}).~Both pressure and temperature obtain same local maximum and minimum values in time at all points, throughout the process, except from the end, after 360 $\mu$s, where the temperature starts an abrupt increase, while the pressure is decreasing due to its periodic behavior.~This temperature increase marks the final stage of the autoignition process.~Among the three points and, despite the fact that the magnitude of oscillation of x$_2$ is not the largest, the autoignition of point x$_2$ first is manifested in Fig. \ref{fig:T2}.~The mechanism by which x$_2$ is autoignited first will be examined in the next sections.

\subsection{Approach}

In order to analyze the dynamics that govern the autoignition process for each one of the three points, the time domain has been divided into periods and the evolution of all quantities will be studied in this context.~The seven periods before autoignition are taken under consideration, for which the points in time that set the periods were calculated according to the pressure profile and are shown in Fig. \ref{fig:periods}.~As starting points of each period there were selected the time points where every pressure pulse generated by the flame front reaches each point.~For example, the times points and periods for the case of x$_1$ are displayed in Table \ref{table:periods}.~It is shown that the periods decrease in time, while the oscillations become more steep, accelerating this way the reactivity of all points.\\[2ex]

	\begin{minipage}[t]{0.5\textwidth}
		\centering
		\includegraphics[scale=0.35]{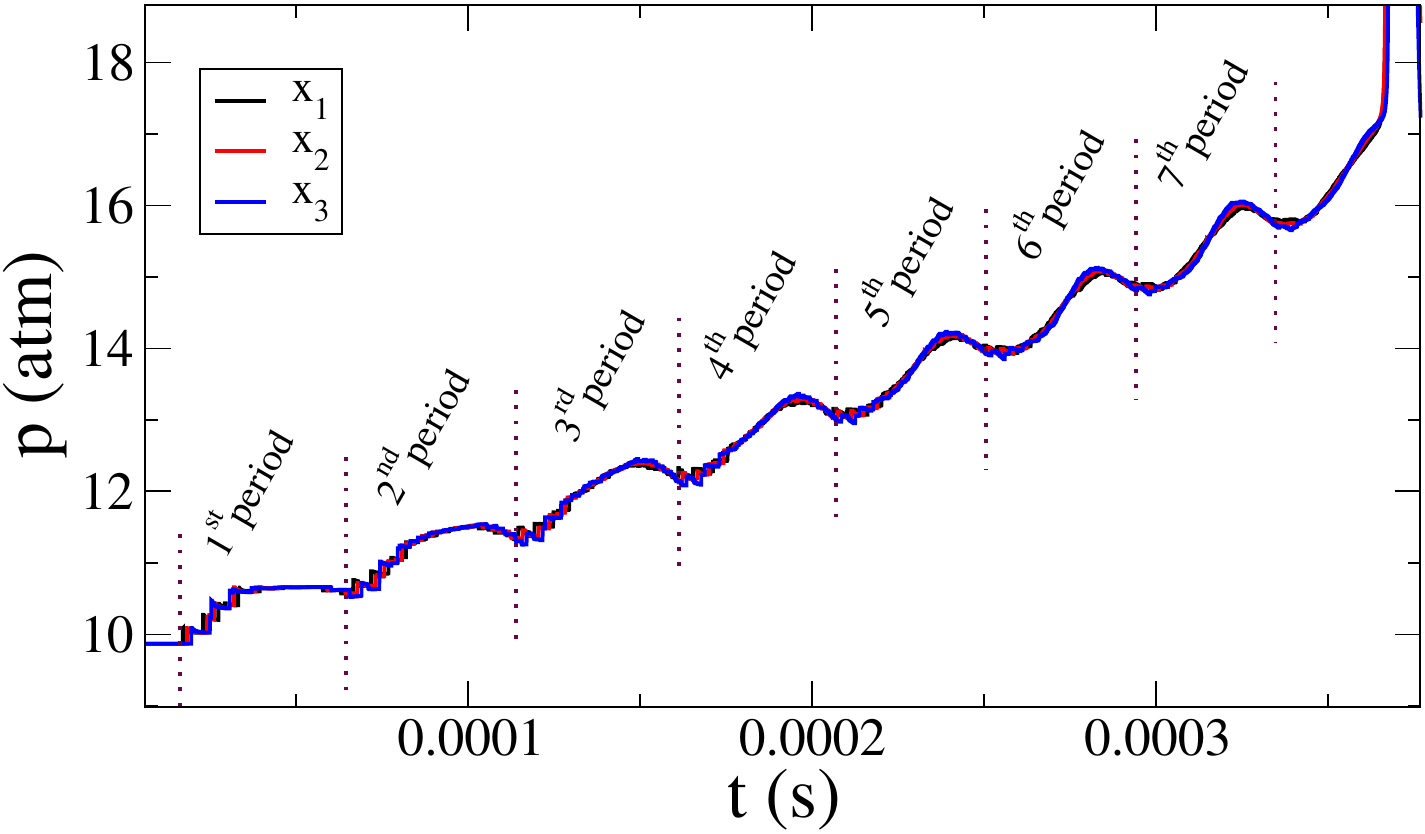}
		\captionof{figure}{Periods of oscillations of x$_1$ on the pressure profile.}
         \label{fig:periods}
	\end{minipage}
\hfill
	\begin{minipage}[b]{0.45\textwidth}
\centering
		\begin{tabular}{c c c c}
		\hline
\rowcolor{Gainsboro!60} 		& t$_{start}$ ($\mu$s)& t$_{end}$ ($\mu$s) & period ($\mu$s) \\
		\midrule		
		1$^{st}$	&	15.17	&	64.58	&	49.40 	\\
		2$^{nd}$	&	64.58	&	113.84	&	49.26 	\\
		3$^{rd}$	&	113.84	&	161.30	&	47.46 	\\
		4$^{th}$	&	161.30	&	206.94	&	45.64 	\\
		5$^{th}$	&	206.94	&	250.65	&	43.71 	\\
		6$^{th}$	&	250.65	&	294.31	&	43.66 	\\
		7$^{th}$	&	294.31	&	334.79	&	40.48	\\
		\hline
\\
		\end{tabular}
	\captionof{table}{Time points and periods of oscillations of x$_1$.}
	\label{table:periods}
	\vfill
\end{minipage}

\begin{figure}[h!]
\centering
\includegraphics[scale=0.34]{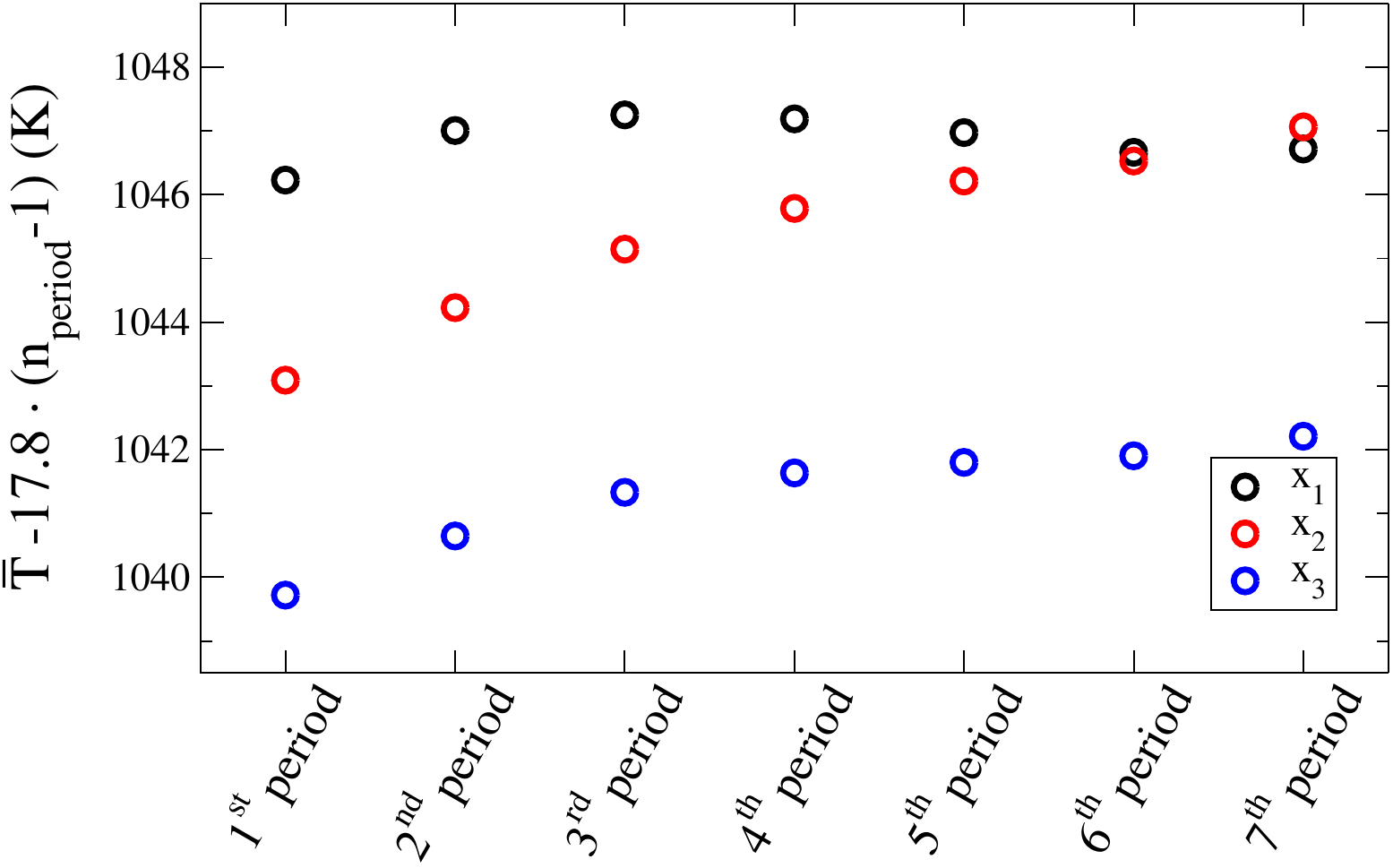} 
\caption{Average temperatures for every period of the oscillations. Values on each period have been reduced by 17.8 K per period, starting from the second period. ~Black: point x$_1$, Red: x$_2$, Blue: x$_3$.}
\label{fig:mean}
\end{figure}

For each period, the average pressure and temperature were calculated.~It was found that the average pressure at each period is pretty much the same for all three cases, while the average temperature at each period is different for every case, as depicted in Fig. \ref{fig:mean}, where the average temperatures have been plotted for every period, by subtracting 17.8 K per period from the values, starting from the second period, in order to scale the differences. This means that values of the first period are displayed correctly, values of second period are -17.8K, values of third period are -35.6K, etc.

~According to Fig.~\ref{fig:mean} the average temperature of each of the points initially differs around 3 K from each other, with $\bar{T}_{x_1}$(black) $>\bar{T}_{x_2}$ (red) $>\bar{T}_{x_3}$ (blue), which is reasonable since they represent points that belong to the region of negative temperature slope.~This difference is being decreased in time, as the average temperature of x$_2$ is increased in a no-linear manner, relatively to the other two which are being subjected to a similar almost linear increase.~Considering that the initial temperature of x$_1$ is greater than that of x$_2$, it is expected that x$_1$'s average temperature of each period to be greater than x$_2$'s everywhere, which is not.~Instead, during the last periods, even though $\bar{T}_{x_3}$ is increased compared to $\bar{T}_{x_1}$ due to the increased magnitude of its oscillation, shortening the difference between $\bar{T}_{x_1}$ and $\bar{T}_{x_3}$ values, the difference between $\bar{T}_{x_2}$ and $\bar{T}_{x_1}$ diminishes until $\bar{T}_{x_2}$ becomes the largest by around 1K.~After the 7$^{th}$ period, towards ignition, this difference becomes even larger, reaching a value of more than 2 K.

The difference of 2K in the average values seems quite small, but considering that the rate coefficients of the reactions that govern the dynamics are an exponential function of temperature, a 1$\%$ perturbation in temperature can alter the values of the rate coefficients by up to 20-25$\%$, which consequently can affect the reaction rates and/or even activate other reactions, which may lead the system to a solution by following different chemical paths.

For a better insight on the effect of the temperature oscillations to the evolution of the system, the explosive timescales were calculated and the reactions that govern the dynamics in every case were studied in the context of CSP tools. Their affection from the small differences on temperature, for every case, were analyzed.~For future reference, the most significant reactions to the dynamics of the system are summarized in Table~\ref{table:reactions}.

\begin{table}[h!]
\centering
\captionof{table}{The most significant reactions to the dynamics of the system.}
\begin{tabular}{r c}
\hline
1f:		& H + O$_2\rightarrow$ O + OH\\
3f:		& O + H$_2\rightarrow$ H + OH\\
4f:		& H$_2$ + OH $\rightarrow$ H$_2$O + H\\
11f:	& H+O$_2$ (+M) $\rightarrow$ HO$_2$ (+M)\\ 
13f:	& HO$_2$ + H $\rightarrow$ OH + OH\\
18f:	& H$_2$O$_2$ (+M) $\rightarrow$ OH + OH (+M)\\
20b:	& H$_2$O$_2$ + H $\leftarrow$ HO2 + H2\\
\hline
\end{tabular}
\label{table:reactions}
\end{table}



\section{CSP Analysis}
\label{CSP}


\subsection{The Algorithmic CSP Tools}
\label{Tools}
Consider the reacting flow (species and energy) governing equations in the ($N+1$)-dimensional vector form:
\begin{equation}
\label{eq:gov1a}
\dfrac{d\bold{z}}{dt} = \bold{L}(\bold{z}) + \bold{g}(\bold{z})
\end{equation}
where $\bold{z} = [\bold{y}, T]^T$ is a column vector that includes the $N$-dimensional column vector of the species mass fractions $\bold{y}$ and the temperature T, the column differential vector operator $\bold{L}(\bold{z})$ represents transport (convection and diffusion) and $\bold{g}(\bold{z})$ represents the chemical kinetics reaction column vector term.~It is assumed that the source term $\bold{g}(\bold{z})$ represents $K$ reversible reactions, whose the forward and reverse reactions are treated separately, so that:
\begin{equation}
\bold{g}=\hat{ \mathbf{S}}_1R^1+\hat{ \mathbf{S}}_2R^2+ \dots +\hat{ \mathbf{S}}_{2K}R^{2K}
\label{ggff2}
\end{equation}
where $\hat{ \mathbf{S}}_k$ and $R^k$ are the generalised stoichiometric vector and rate of the k-th (k=1,2K) unidirectional reaction, respectively \citep{Valorani2003,Diamantis2015b}.~In CSP form Eq.~(\ref{eq:gov1a}) yields:
\begin{equation}
\label{eq:gov2a}
\dfrac{d\bold{z}}{dt} = \sum_{n=1}^{N+1}\bold{a}_nf^n         \qquad \qquad \qquad     f^n=\bold{b}^n .~\left[  \bold{L}(\bold{z}) + \bold{g}(\bold{z})   \right]
\end{equation}
where $\bold{a}_n$ is the $(N+1)$-dimensional~CSP column basis vector of the $n$-th mode and $f^n$ is the related amplitude and $\bold{b}^n$ is the corresponding $(N+1)$-dim.~row n-th dual basis vector ($\bold{b}^i \cdot \bold{a}_j = \delta_j^i$)  \citep{Lam1988,Lam1994,hadjinicolaou1998asymptotic}; $f^n$ is always set positive, with the appropriate choice of the sign of $\bold{b}^n$ (and of $\bold{a}_n$ in order to preserve orthogonality).~The n-th CSP mode $\bold{a}_nf^n$ in Eq.~(\ref{eq:gov2a}) is characterised by (i) the time scale $\tau_n$, which measures the time frame of its action, (ii) the amplitude $f^n$, which measures the impact of its action and (iii) the variables that relate to this mode.~When the $M$ fastest time scales ($\tau_1<\dots<\tau_M$)  of the system in Eq.~(\ref{eq:gov2a}) are exhausted, the system reduces to:
\begin{equation}
\label{eq:gov3a}
f^m\approx 0\quad(m=1, \dots M) \qquad \qquad \qquad \dfrac{d\bold{z}}{dt} \approx \sum_{n=M+1}^{N+1}\bold{a}_nf^n \ \
\end{equation}
The $M$-dimensional algebraic system $f^m\approx 0$ defines an approximation of the low dimensional slow invariant manifold (SIM) on which the solution evolves and the $(N+1)$-dimensional system of ODEs in Eq.~(\ref{eq:gov3a}) governs the flow on the SIM, which is characterised by the fastest of the slow time scales, when the solution evolves sufficiently far from the boundaries of the SIM \citep{Goussis2013PD,Maris2015PD}.

In reactive processes, the fastest time scales in the dynamics of the system in Eq.~(\ref{eq:gov1a}) usually originate from the chemical kinetics term $\bold{g}(\bold{z})$ \citep{Valorani2003,najm2009analysis,Lu2010,goussis2005reactive,Luo2012265,Shan2012b,kooshkbaghi2015n,tingas2015autoignition}.~Considering the case where the $n$-th nonzero eigenvalue ${\lambda}_n$ of the Jacobian $\bold{J}$ of $\bold{g}(\bold{z})$ is real (the extension to complex pairs is straightforward \citep{Goussis2006}), the time scales introduced by the chemical kinetics term are approximated by the relation $\tau_n = |\lambda_n|^{-1}$;  $n=1, \dots N-E+1$, where $E$ is the number of elements in the chemical kinetics mechanism employed.~When $\lambda_n$ is positive (negative), the related time scale $\tau_n$ is an explosive (dissipative) one, since it relates to components of the system that tend to deviate from (approach) equilibrium.~The eigenvalue is defined as ${\lambda}_n = \boldsymbol{\beta}^n \cdot \bold{J} \cdot  \boldsymbol{\alpha}_n$, where $\boldsymbol{\alpha}_n$ and $ \boldsymbol{\beta}^n$ are the $n$-th right (column) and left (row) eigenvectors of $\mathbf{J}$, respectively.~The $n$-th eigenvalue can be expressed as:
\begin{equation}
\label{eq:gov6}
{\lambda}_n =  \boldsymbol{\beta}^n  \bold{J} \boldsymbol{\alpha}_n  = \boldsymbol{\beta}^n  \cdot  \sum_{k=1}^{2K}  grad \left( \hat{ \mathbf{S}}_kR^k \right)  \cdot  \boldsymbol{\alpha}_n = c_1^n + ... + c_{2K}^n
\end{equation}
since according to Eq.~(\ref{ggff2}), $\bold{J}=grad(\hat{ \mathbf{S}}_1R^1)+ \dots +grad(\hat{ \mathbf{S}}_{2K}R^{2K})$ \citep{Goussis2005a,Diamantis2015b}.~The expression in Eq.~(\ref{eq:gov6}) suggests the introduction of the {\em Time scale Participation Index} (TPI):
\begin{equation}
\label{eq:gov7}
J^n_k = \dfrac{c_k^n}{|c_1^n| + ... + |c_{2K}^n|} \qquad \qquad \qquad  \qquad \sum_{k=1}^{2K}|J_k^n|=1
\end{equation}
where $n=1,\dots (N-E+1)$, $k=1, \dots 2K$ \citep{Goussis2005a}.~$J^n_k$ measures the relative contribution of the $k$-th reaction to  ${\lambda}_n$ and, therefore, to $\tau_n$.~A positive (negative) $J_k^n$ implies that the $k$-th reaction contributes to the explosive (dissipative) character of the $n$-th time scale $\tau_n$.~TPI has been used successfully in a large number of autoignition problems in identifying the chemical reactions that determine the characteristic time scale \citep{valorani2020computational1, valorani2020computational2} and thus influence the ignition delay time \citep{manias2015initiation, manias2015algorithmic, tingas2018ch4, khalil2019algorithmic, manias2021nh3, rabbani2021ch3oh}.

Considering the n-th mode, let us denote the sum of all positive and all negative $J^n_k$ as $\sum_+^{e,f}$  and $\sum_-^{e,f}$, where $\sum_+^{e,f}-\sum_-^{e,f}=1$ \citep{Diamantis2015b}.~Clearly, the magnitude of these two quantities assess the relative activity of the reactions that promote the explosive character of $\tau_n$ ($\sum_+^{e,f}$) and of those that promote its dissipative character ($\sum_-^{e,f}$).~For an explosive time scale, $\sum_+^{e,f}>-\sum_-^{e,f}$.

Given the expression in Eq.~(\ref{ggff2}) for the chemical kinetics term $\bold{g}$, the amplitude of the n-th CSP mode $f^i$ ($i=1,\dots,N+1)$ can be expressed as:
\begin{equation}
f^n \ = \ f^n_1 + ... + f^n_{2K}+ f^n_{2K+1}
\label{ggff1}
\end{equation}
where $f^n_k=({\bf b}^n \cdot \hat{ \mathbf{S}}_k)R^k$ (k=1,2K) is the contribution of the $k$-th reaction to the amplitude $f^n$ and $f^n_{2K+1}=({\bf b}^n \cdot \mathbf{L})$  is the contribution of the transport term.~The contribution of each reaction and of the transport term to $f^n$ can be assessed by the \emph{Amplitude Participation Index} (API):
\begin{equation}
P^n_k = \frac{f^n_k({\bf y})}{\sum_{j=1}^{2K+1}|f^n_j({\bf y})| }  \qquad \qquad \qquad  \qquad \sum_{k=1}^{2K+1}|P_n^k|=1
\label{PI}
\end{equation}
where $n=1,N+1$ and $k=1,...,2K+1$ \citep{Goussis1992,Lam1994,Valorani2003,Najm2010,Prager2011,gupta2011classification,Diamantis2015b,kooshkbaghi2015n}.~Since $f^n>0$, positive (negative) values of $P^n_k$ indicate a contribution of the k-th reaction towards strengthening (weakening) the impact of the n-th mode.~In the case of an exhausted mode $f^m\approx 0$ (m=1,M) (see Eq.~(\ref{eq:gov3a})), $P_1^m + . . . +P_{2K}^m \approx 0$.~Therefore, a relatively large value of $P_k^n$ indicates a large contribution from the $k$-th reaction to (i) the cancellations between the various terms in the expression in Eq.~(\ref{ggff1}) and (ii) the m-th component of the SIM, the development of which is characterised by the $m$-th fast time scale, $\tau_m$. API has been used successfully in a large number of highly turbulent flames problems in identifying the impact of the dominant chemical reactions to the system \citep{manias2018analysis, manias2019investigation, manias2019dynamics, manias2019topological}.

The variables with the greatest influence on the $m$-th fast time scales are identified by the \emph{CSP Pointer} (PO) for the $m$-th mode:
\begin{equation}
{\bf D}^m=diag[{\bf a}_m{\bf b}^m]       =  [a^1_mb^m_1, ... , a^N_mb^m_N]^T
\label{PO}
\end{equation}
where due to orthogonality $a^1_mb^m_1+ ... +a^N_mb^m_N = 1$, ($m=1, \dots ,M$) \citep{Goussis1992,Lam1994,Valorani2003}.~A large value of $a^j_mb^m_j$ ($j=1,...,N$) denotes a strong relationship between the $j$-th variable and the $m$-th \emph{CSP mode} ${\bf a}_m f^m$ and the related fast time scale $\tau_m$. The use of CSP Pointer has been successfully employed in previous works \citep{manias2016mechanism, khalil2021no, rabbani2022dominant, manias2022effect} as an indicator of probable additives in the starting fuel/air mixture from the set of intermediate species.

The CSP methodology employed was introduced in 1989 \citep{Lam1988} and it has since been studied, employed, and verified extensively in a large number of combustion, biological and  pharmacokinetics problems in accurately identifying the dominant chemical processes \citep{kooshkbaghi2015n,tingas2015autoignition,Goussis2006,Goussis2005a,Diamantis2015b,Habib2006,diamantis2009two,Kourdis2013b,Shan2014,diamantis2015reactions,manias2016mechanism,tingas2016ignition,tingas2016comparative,tingas2016algorithmic,patsatzis2016asymptotic}.



\subsection{CSP Diagnostics}
\label{CSP_Diagnostics}

\subsubsection*{The developing explosive timescales}

For the case considered, two explosive time scales develop; the fast explosive time scale, say $\tau_{e,f}$, which exists from the beginning and acts throughout the process and the slow explosive time scale, say $\tau_{e,s}$, which appears after 190 $\mu$s, for all three points under study.~Both time scales act until the point where the temperature reaches the middle of its steep rise, after 366 $\mu$s for all points, which marks the ignition delay time. The period from the beginning of the process until the steep rise of temperature is called the \textit{explosive stage} of the process.

\begin{figure}[h]
\hspace{-5mm}
\begin{subfigure}[b]{0.31\textwidth} 
\includegraphics[scale=0.22]{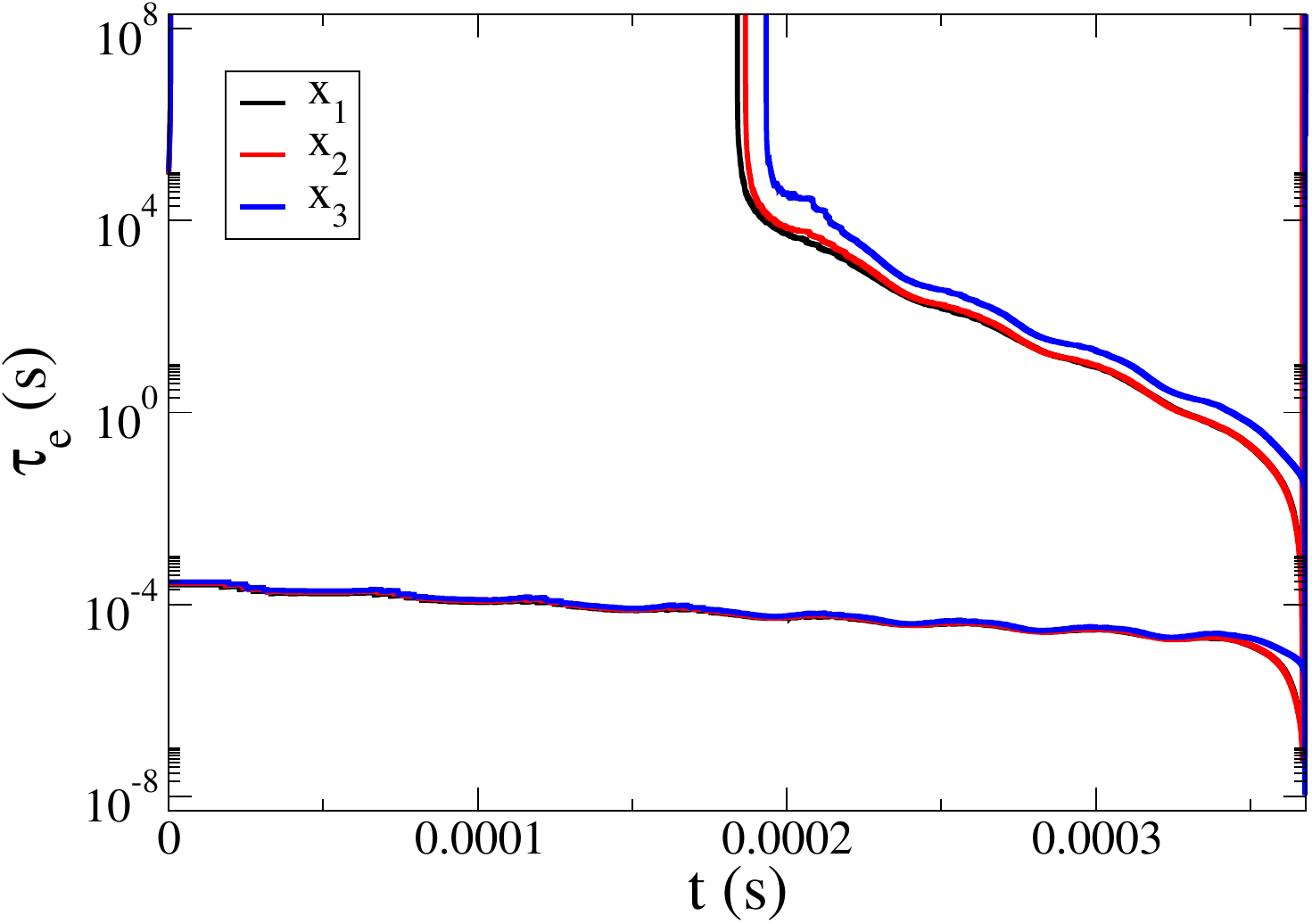}
\caption{}
\label{fig:exp1}
\end{subfigure}
\hfill
\begin{subfigure}[b]{0.31\textwidth} 
\includegraphics[scale=0.22]{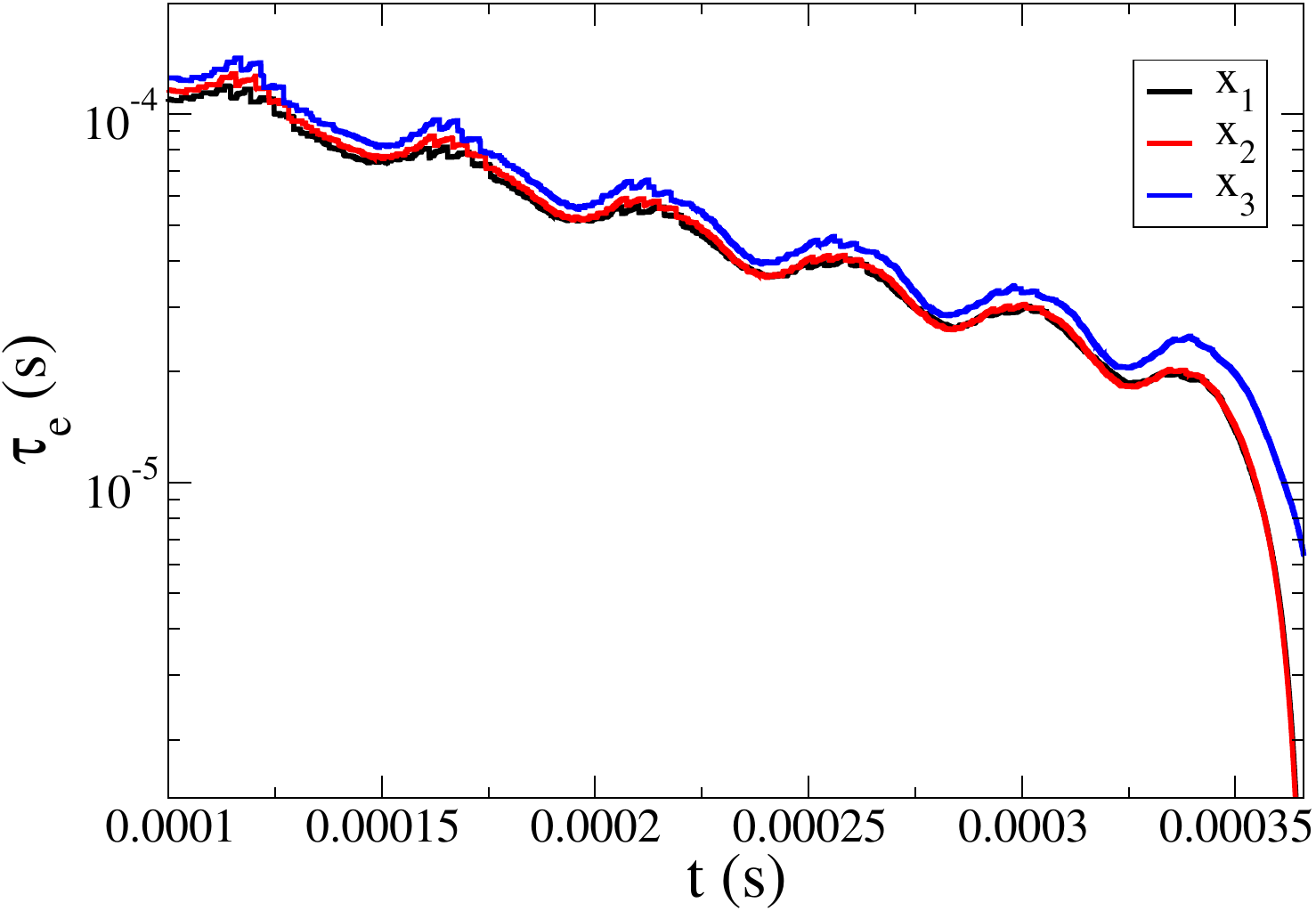}
\caption{}
\label{fig:exp2}
\end{subfigure}
\hfill
\begin{subfigure}[b]{0.31\textwidth} 
\includegraphics[scale=0.22]{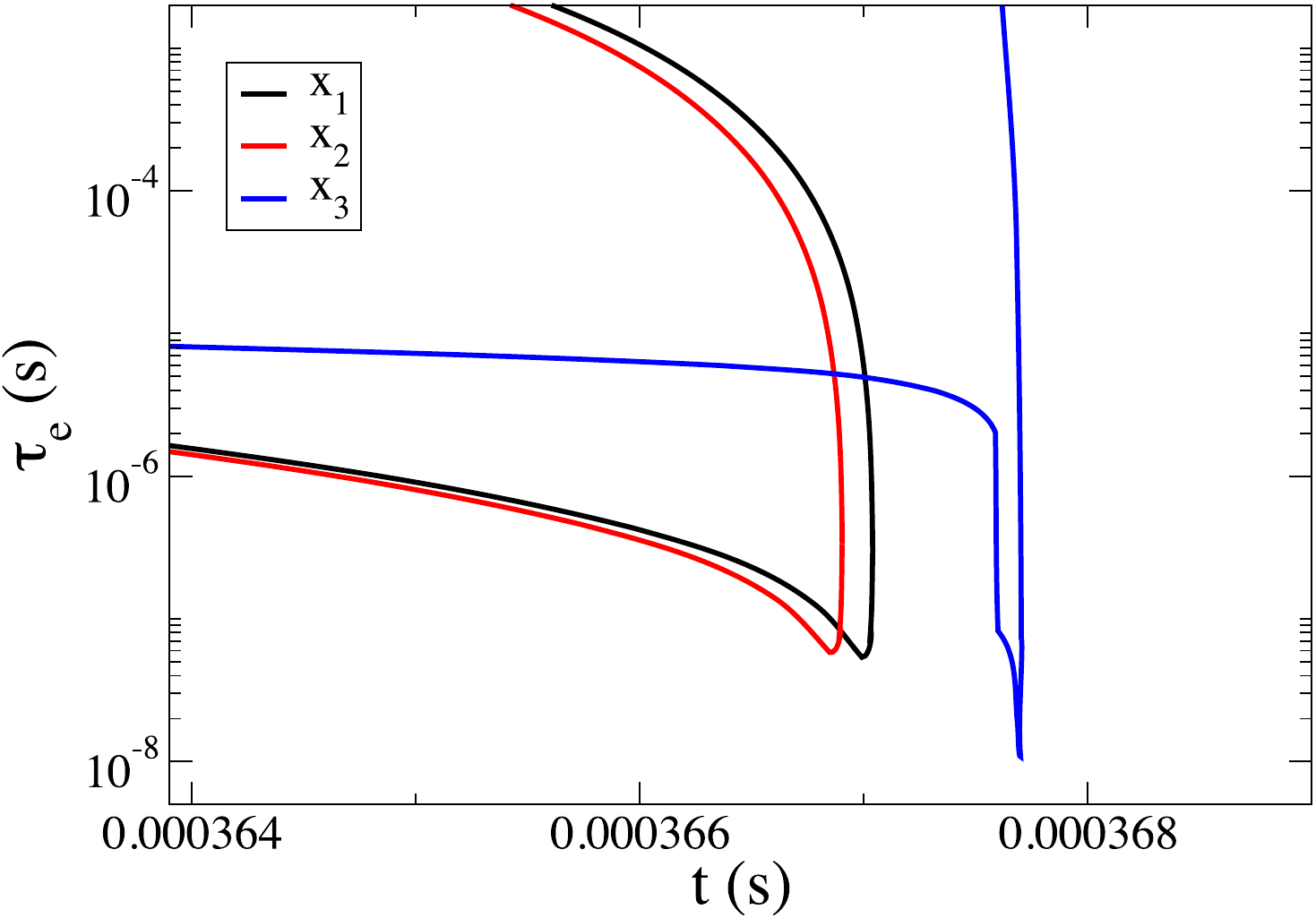}
\caption{}
\label{fig:exp3}
\end{subfigure}
\hfill
\caption{The evolution of the explosive time scales for the 3 tpoints under investigation. Black: x$_1$, red: x$_2$ and blue: x$_3$.}
\label{fig:exp}
\end{figure}

Figure \ref{fig:exp1} shows the development of the fast and slow explosive time scales for the three points under investigation.~It is shown that, in all three cases, the fast explosive time scale is much faster than the slow one throughout the duration of the explosive stage. Thus, it is the one that characterizes the dynamics during this stage.~For all cases, $\tau_{e,f}$ is accelerating in a steady low rate, following the same behavior of pressure's, oscillating in the same periods, as indicated in Fig. \ref{fig:exp2}.~Despite its oscillating behavior, the magnitude of $\tau_{e,f}$ does not change significantly, throughout the largest part of the explosive stage.~For all cases, the magnitude of $\tau_{e,f}$ is around $\sim$4.5$\times$10$^{-4}$ at 0 and accelerates to $\sim$2.5$\times$10$^{-5}$ at $\sim$350 $\mu$s, where a rapid acceleration is observed, followed by a more rapid deceleration, until it meets the slow time scale, which is also rapidly accelerating in this final part of the explosive stage, as indicted in Fig.~\ref{fig:exp3}.

The average $\tau_{e,f}$ for each oscillation were calculated for the three cases and it was found that the average $\tau_{e,f}$ of x$_2$ becomes the fastest at around $\sim$300 $\mu$s, indicating that faster dynamics drive the evolution of the x$_2$ case.~As a result, ignition takes place first at x$_2$, as it is also indicated by the earlier occurrence of the minimum value of $\tau_{e,f}$.~The $\tau_{e,f}$ of x$_3$ is found to be slower than the other two cases everywhere as a result of the relative less reactive state.~Additionally, the ignition at x$_2$ manifests gradually, following a gradual ignition at x$_1$, while at x$_3$ it manifests abruptly.~The difference in the intensity of ignition correlates perfectly with the difference in the minimum values of $\tau_{e,f}$ exhibited towards the end of the explosive stage.~These minimum values are $\tau_{e,f}$ = 6$\times$10$^{-8}$ for the first two cases and $\tau_{e,f}$ = 10$^{-8}$ for the third.~Furthermore, the profile of $\tau_{e,f}$ at Fig.~\ref{fig:exp3} of x$_2$ indicates that the abrupt acceleration of $\tau_{e,f}$ at the end of its explosive stage is caused by the pressure pulse generated by the autoignion at x$_2$, so it is expected that the processes that dominate is evolution to differ from the ones of x$_2$ and x$_1$.~In order to obtain an insight of the dominant processes of each case, the CSP diagnostic tools were used and the reactions and species that affect the most the explosive dynamics of each case were identified and are presented next.
\subsubsection*{CSP TPI}

\begin{figure}[h!]
\begin{subfigure}[b]{0.45\textwidth} 
\centering
\includegraphics[scale=0.25]{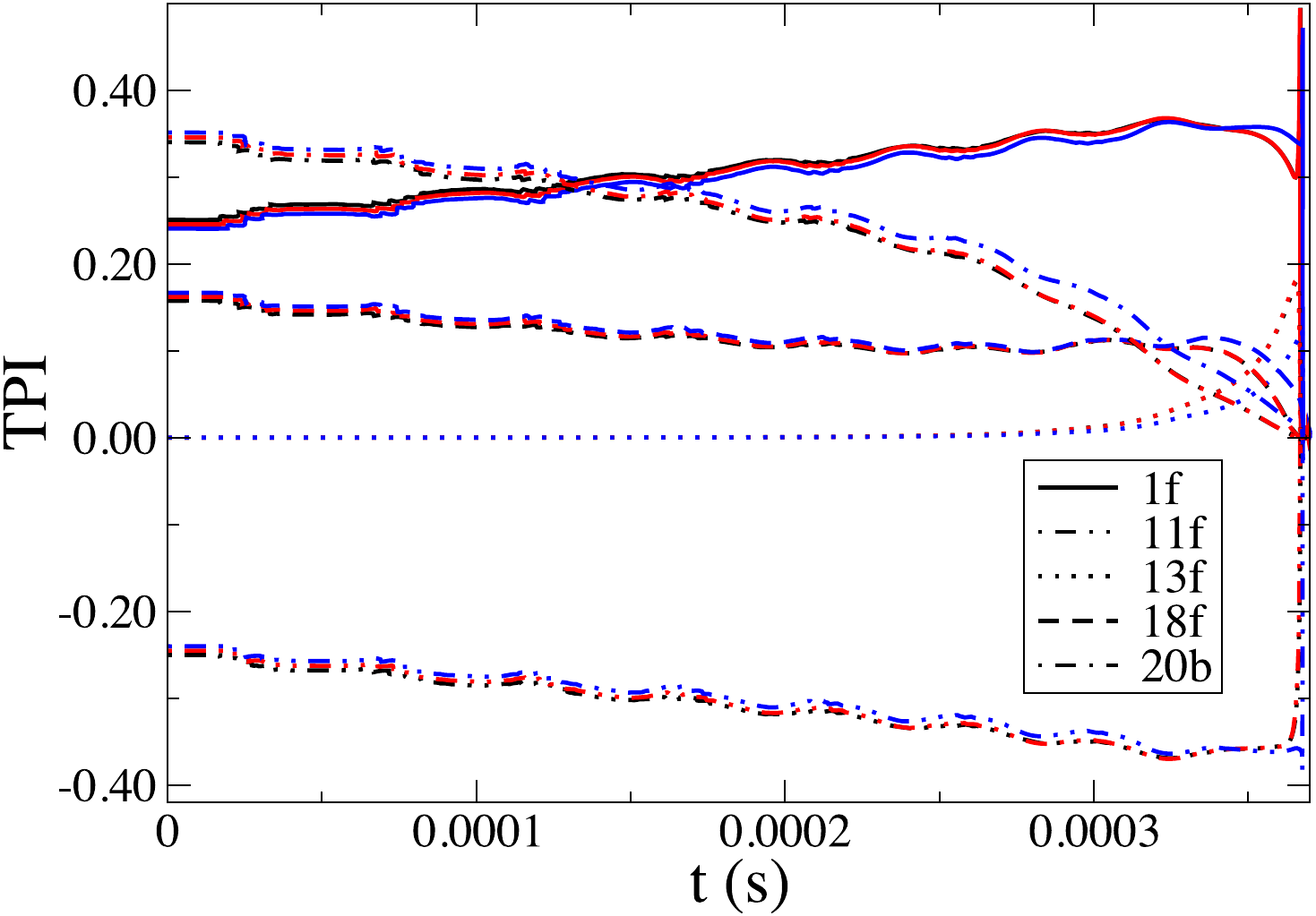}
\caption{}
\label{fig:TPI1}
\end{subfigure}
\hfill
\begin{subfigure}[b]{0.45\textwidth} 
\centering
\includegraphics[scale=0.25]{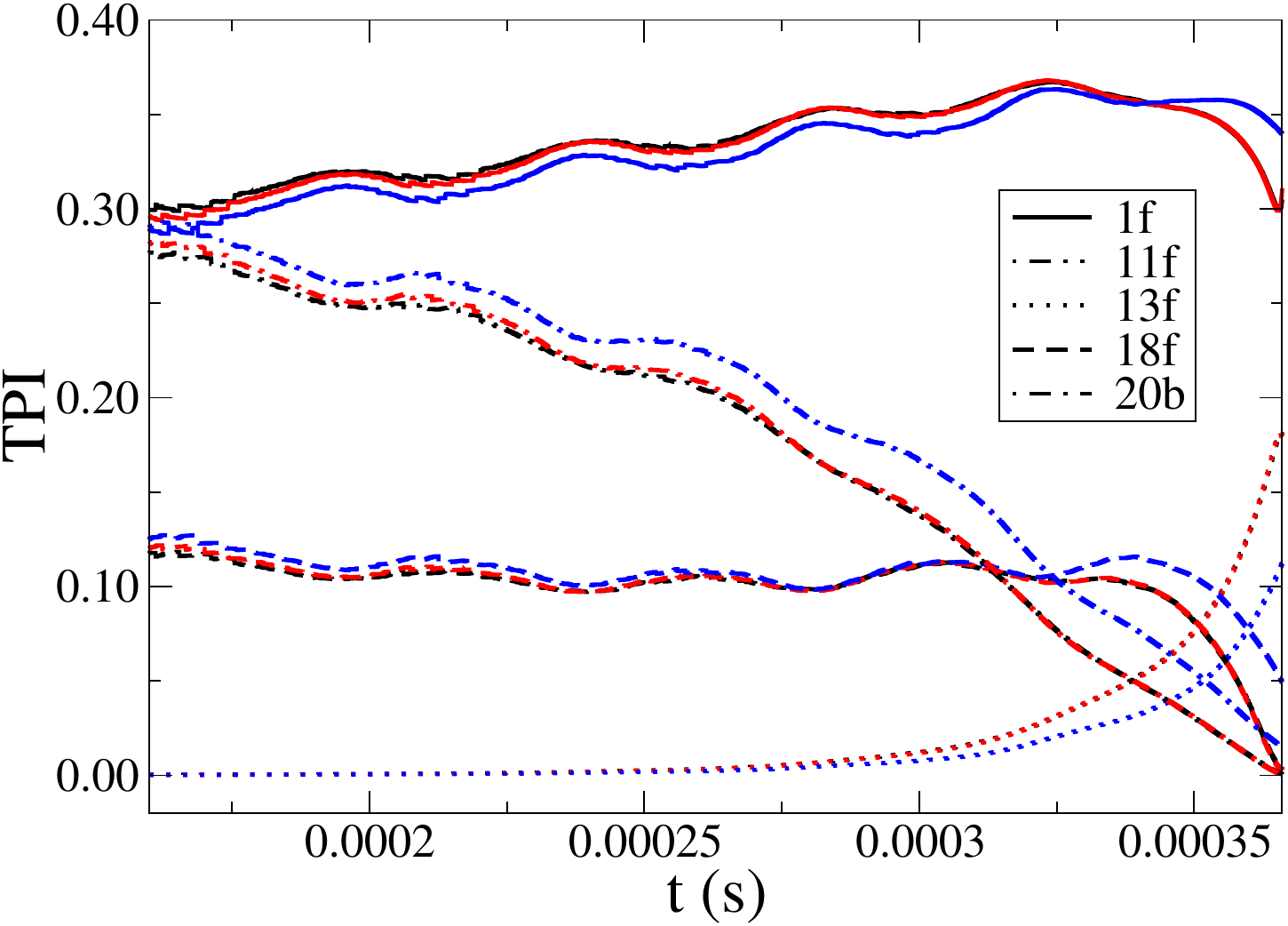}
\caption{}
\label{fig:TPI2}
\end{subfigure}\hfill\hfill
\\
\hfill\hfill
\begin{subfigure}[b]{0.45\textwidth} 
\centering
\includegraphics[scale=0.25]{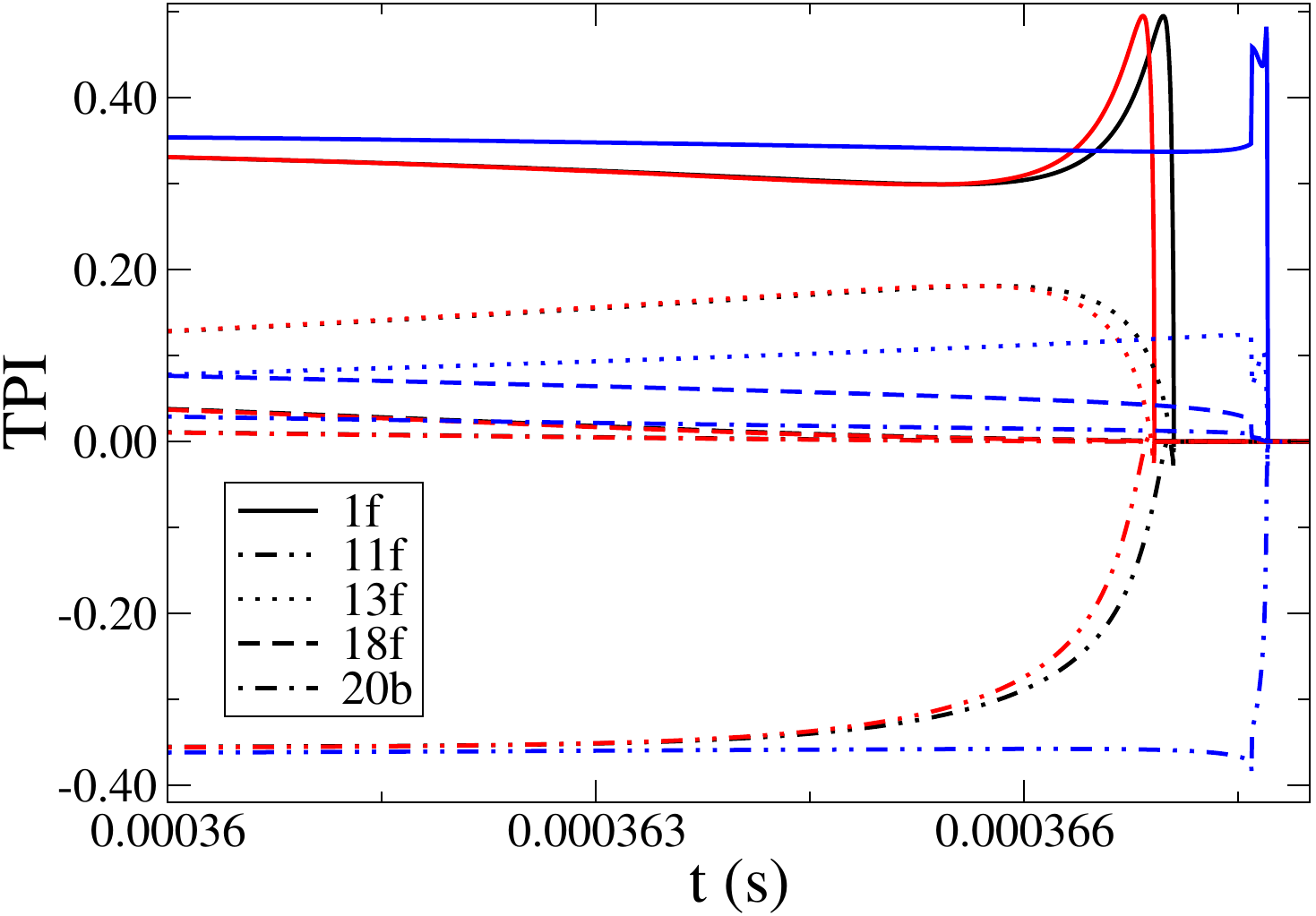}
\caption{}
\label{fig:TPI3}
\end{subfigure}
\hfill
\begin{subfigure}[b]{0.45\textwidth} 
\centering
\includegraphics[scale=0.25]{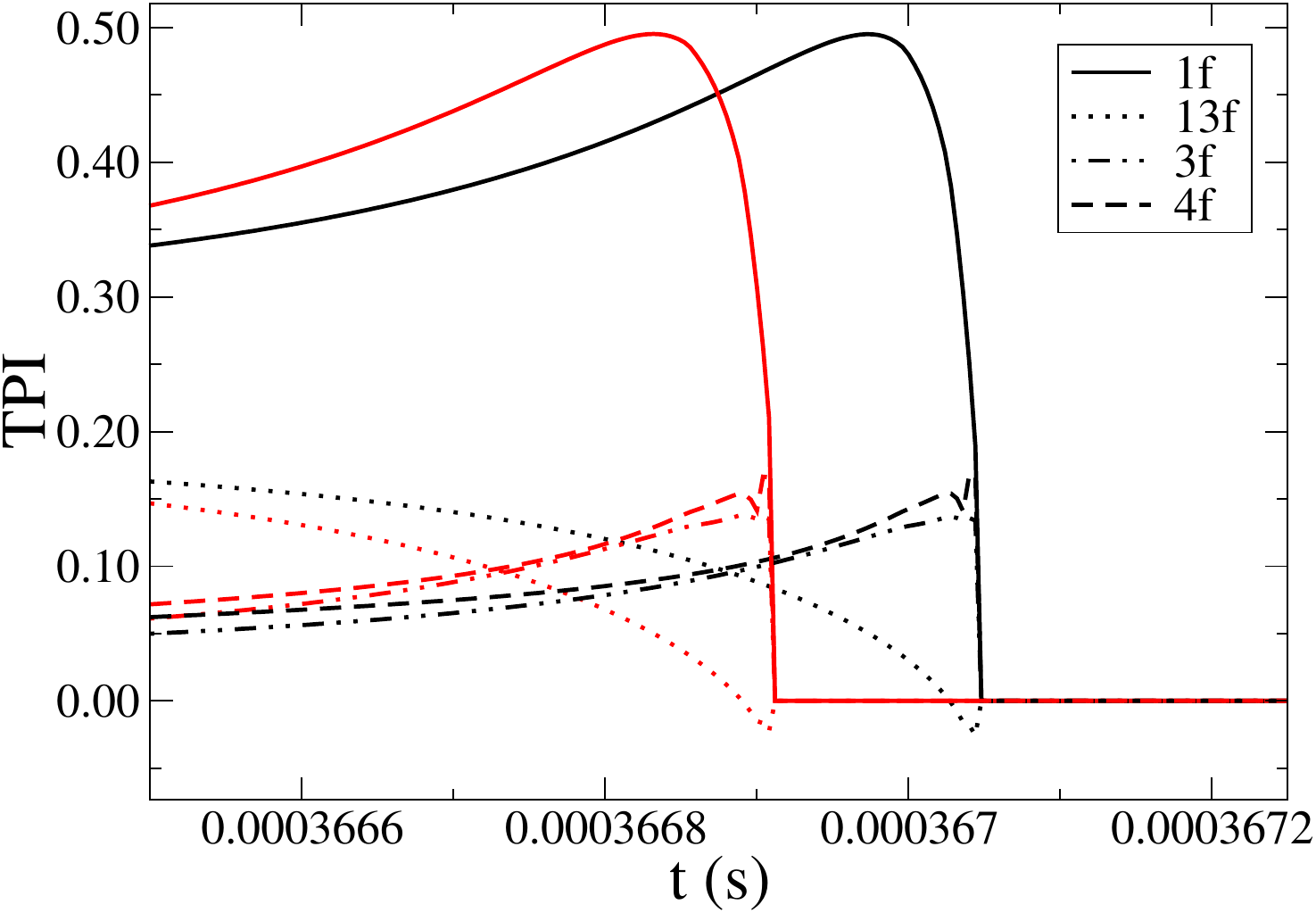}
\caption{}
\label{fig:TPI4}
\end{subfigure}\hfill\hfill
\caption{The evolution of the largest values of TPI, for the 3 tpoints under investigation. Black: x$_1$, red: x$_2$ and blue: x$_3$.}
\label{fig:TPI}
\end{figure}

The reactions that affect the most the fast explosive time scale, $\tau_{e,f}$, which is the characteristic time scale, were calculated by using the \textit{CSP Timescale Participation Index}.~The way of which these reactions are promoting or opposing the explosive character of  $\tau_{e,f}$, thus, affecting the dynamics that govern the process, was quantified and the results are depicted in Fig.~\ref{fig:TPI}.\\


First the diagnostics were obtained in a few points close to the ignition delay time.~It was found that the actions of the reactions that affect the most $\tau_{e,f}$ close to ignition, was starting from the beginning of the process, as shown in Fig.~\ref{fig:TPI1}.~In particular, it is shown that, initially and for all three cases, the major contributor to $\tau_{e,f}$ is reaction 20b: H$_2$O$_2$ + H $\leftarrow$ HO2 + H2, supporting the explosive character by $\sim$35$\%$ at 0$\mu$s and gradually dropping to zero until the ignition time.~The role of this reaction is to generate the necessary H radical pool that is needed in the final part of the \textit{explosive stage}, along with the H$_2$O$_2$ species that has been found to be a very important species of the autoignition process.

For all cases, around 150 $\mu$s, the major contributor to $\tau_{e,f}$ becomes the chain branching reaction 1f: H+O$_2\rightarrow$ O+OH.~This reaction has been found to  support the explosive character of $\tau_{e,f}$ throughout the whole time interval.~As a result, reaction 1f is the one that set the timeframe of action of the explosive mode and throughout the process is the reaction that provides with OH radicals the system.~Beginning form the value 0.25 at 0$\mu$s, its relative contribution gradually increases to around 0.35 at 320 $\mu$s (Fig.\ref{fig:TPI2}).~Then, after a small decrease, at $\sim$365 $\mu$s, its relative contribution rises up to 0.5, right before the ignition occurs and then it falls to zero (Fig.\ref{fig:TPI3}). 

Reaction 18f: H$_2$O$_2$ (+M) $\rightarrow$ OH + OH (+M) is the third in value contributor to the explosive character of the timescale, promoting its explosive character, as well, in all three cases. Starting from 0.18 at 0 $\mu$s, its contribution decreases steadily to 0.13 at $\sim$350 $\mu$s and, after that, the decrease becomes more rapid, until the contribution reaches 0.~Reaction 18f is a third body reaction that provides with radicals the OH pool, so by the end of the explosive stage, its relative contribution to the explosive dynamics vanishes.~Last, reaction 13f: HO$_2$ + H $\rightarrow$ OH + OH, is the last of the major positive contributors to $\tau_{e,f}$, but only around the end of the explosive stage, from $\sim$ 300 $\mu$s, which increases to reach 0.2 at $\sim$366 $\mu$s, before it rapidly drops to 0, as result of the competition with reaction 11f: H+O$_2$ (+M) $\rightarrow$ HO$_2$ (+M), which produces the HO$_2$ radical consumed by 13f.~Reaction 11f is the major contributor that opposes the explosive character of the $\tau_{e,f}$.~Its relative negative contribution is the opposite of reaction 1f. This means that it competes with the action of reaction 1f, which is reasonable, since they both consume of H and O$_2$.~Finally, Fig.~\ref{fig:TPI4} demonstrates that right before ignition occurs, when the system exhibits its faster dynamics, for all cases, reactions 3f and 4f appear and contribute to a lesser but also significant degree to the generation of $\tau_{e,f}$.~Reaction 4f describes the chemical path through which the explosive stage is terminated by consuming the OH radicals.

In general, the TPI relative contributions of all points can be grouped in three stages:~First stage, form 0 $\mu$s to $\sim$350 $\mu$s, where all relative contributions evolve slowly and no rapid changes are observed, as seen in Fig. \ref{fig:TPI1}. The second stage, which is manifested in Fig. \ref{fig:TPI2}, lasts from $\sim$350 $\mu$s to 364 $\mu$s, where the behavior of the contribution changes; some start to decrease, others start to increase. Third, after 364 $\mu$s, rapid changes occur in the values of the relative contributions, until they all reach to zero, at the end of the explosive stage. This is indicated in Fig. \ref{fig:TPI3}.~During the two first stages, no significant differences were found between the three cases.~However, during the third stage, and more precisely at around $\sim$365 $\mu$s, the relative contribution of reaction 1f of case 2 starts to increase, while the one of 11f decreases, in contrast to the other two cases, were their respective contributions remain constant.~Only the contributions of x$_1$ follow after a very small delay, as a result of being the neighbor point with the highest reactivity.~Nevertheless, the respective change is not evident in the fast explosive timescale profiles (Fig.\ref{fig:exp3}).

\begin{figure}[h!]
\centering
\includegraphics[scale=0.3]{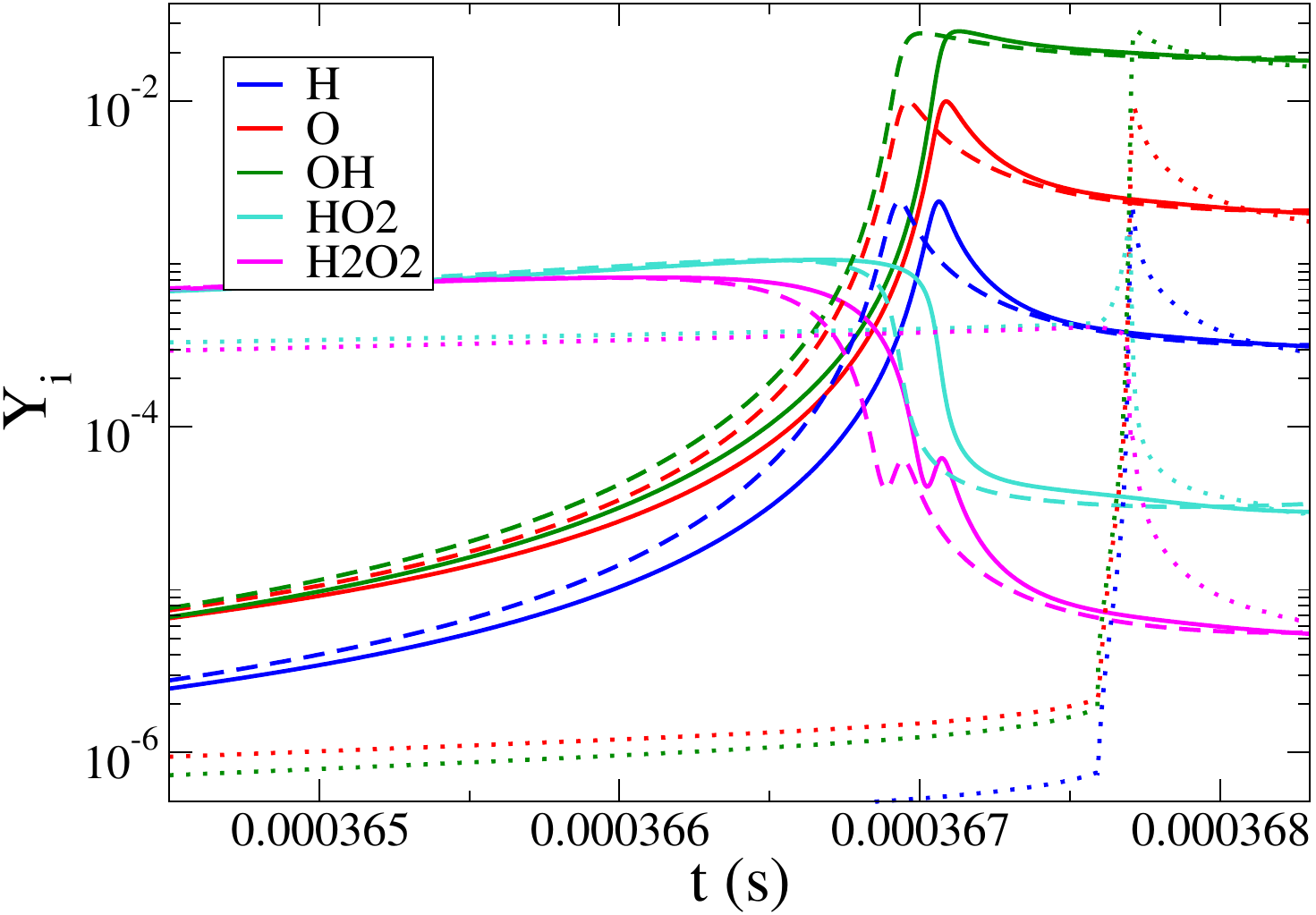} 
\caption{Evolution of the mass fractions of selected species at the end of the explosive stage, for the three cases. ~Solid: x$_1$, dashed: x$_2$, dotted: x$_3$.}
\label{fig:species}
\end{figure}

For that period it was found that the respective to case 2 radical pool is bigger compared to the others.~Figure~\ref{fig:species} displays the evolution of the mass fractions of selected species during the last part of the explosive stage.~It is shown that the concentration of H,O and especially OH radical pools are bigger for case 2, as a result of its faster explosive dynamics, expressed mainly by reaction 1f.~Furthermore, the concentration of H$_2$O$_2$ and HO$_2$ species drops faster at case 2 as a result of the reduced action of reactions 11f  and 13f.~In order to identify how these specific species or if even others are related to the fast explosive mode that drives the system, the   \textit{CSP Pointer tool} was used and the results are disgusted next.

\subsubsection*{CSP Pointer}

Fig. \ref{fig:Po} shows the variables that are most related to the explosive mode, for all three points under consideration.~The variables that were identified to be most related to the explosive mode are all important radicals and species to combustion of hydrogen and hydrocarbon fuels.~Furthermore, close to the end of the explosive stage, temperature was found to be of most importance, for all cases.~Fig. \ref{fig:Po} indicates that the evolution of these pointed variables, can be grouped in three stages:

\begin{figure}[h!]
\begin{subfigure}[b]{0.45\textwidth} 
\centering
\includegraphics[scale=0.25]{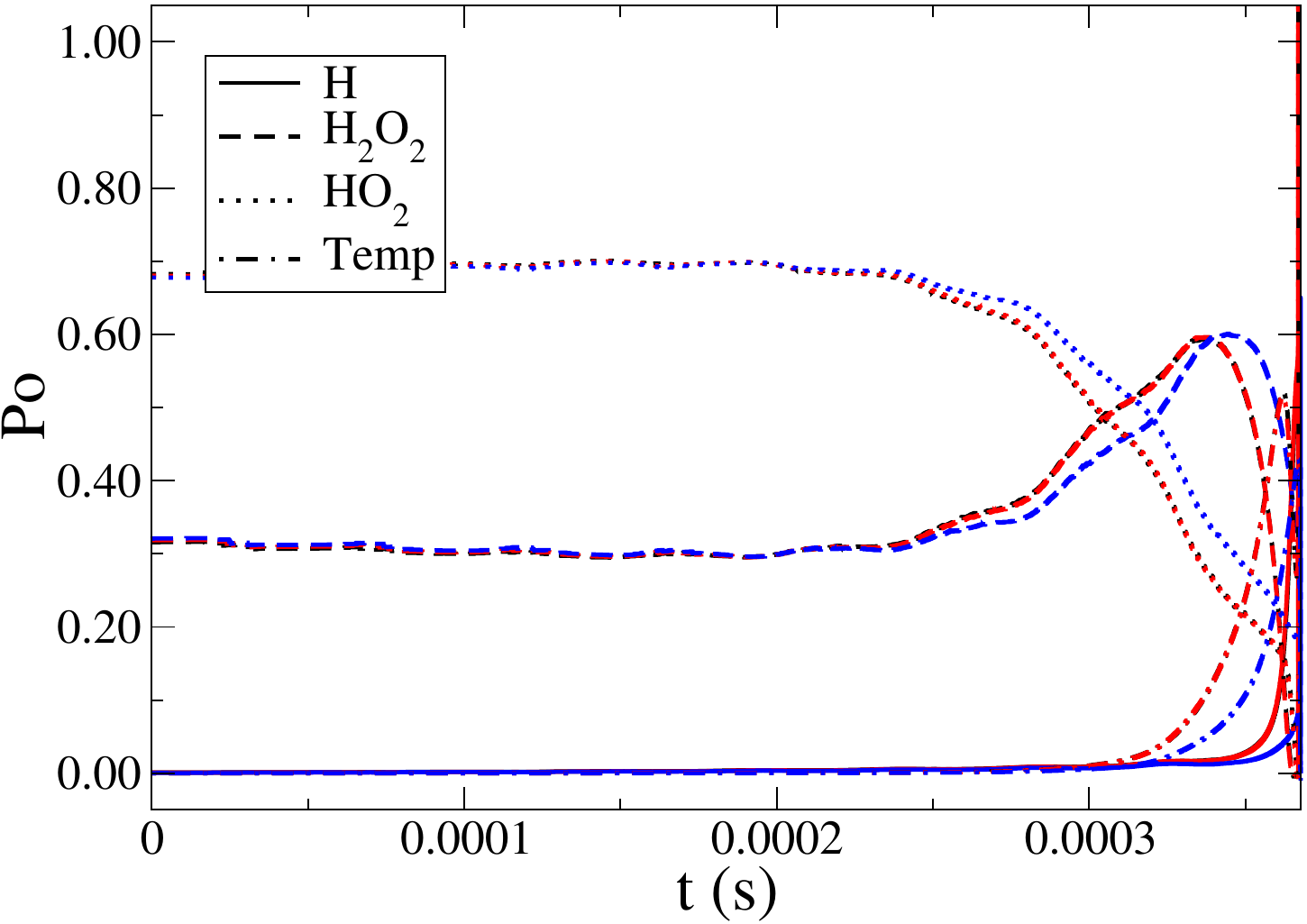}
\caption{}
\label{fig:Po1}
\end{subfigure}
\hfill
\begin{subfigure}[b]{0.45\textwidth} 
\centering
\includegraphics[scale=0.25]{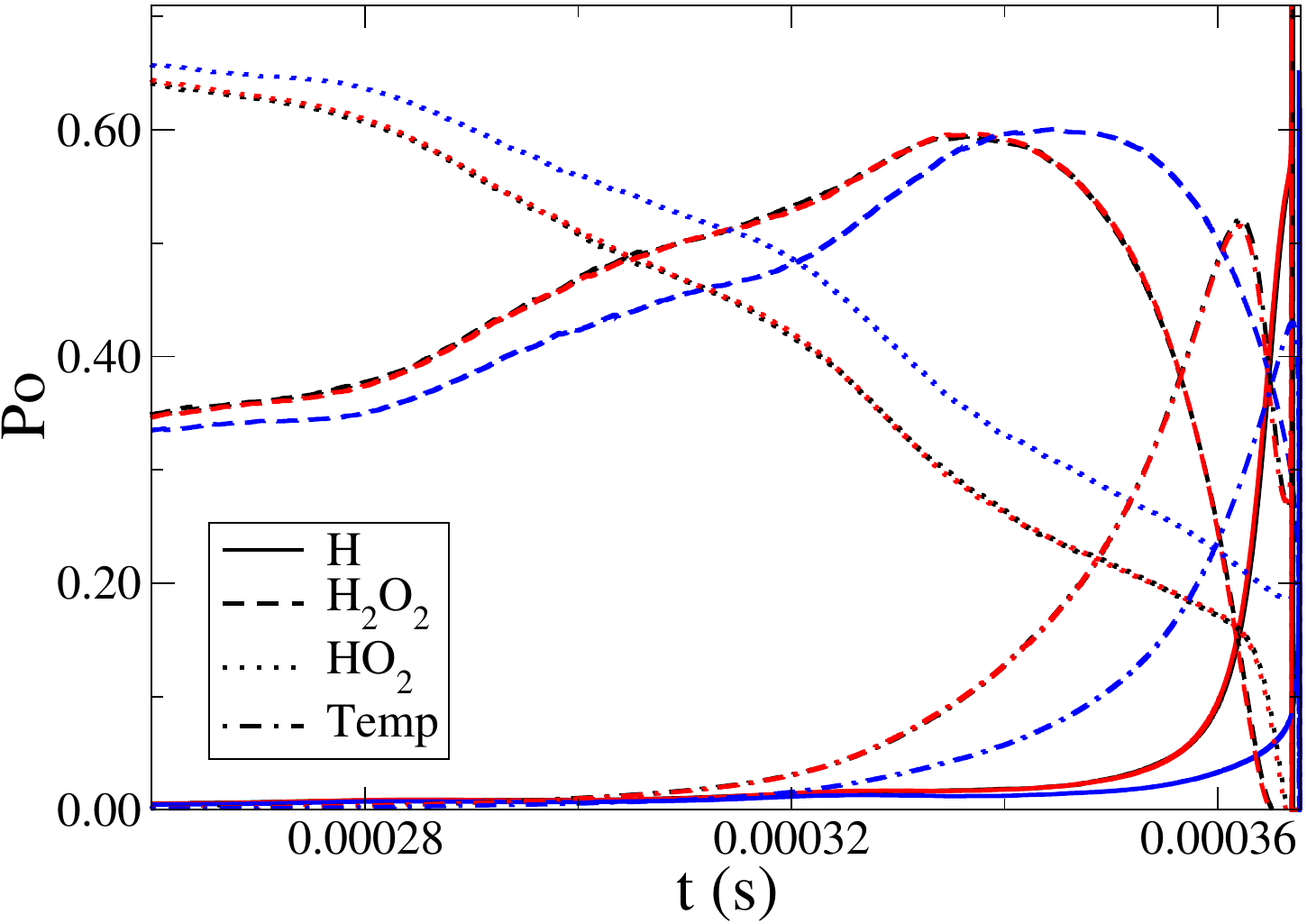}
\caption{}
\label{fig:Po2}
\end{subfigure}\hfill\hfill
\\
\hfill\hfill
\begin{subfigure}[b]{0.45\textwidth} 
\centering
\includegraphics[scale=0.25]{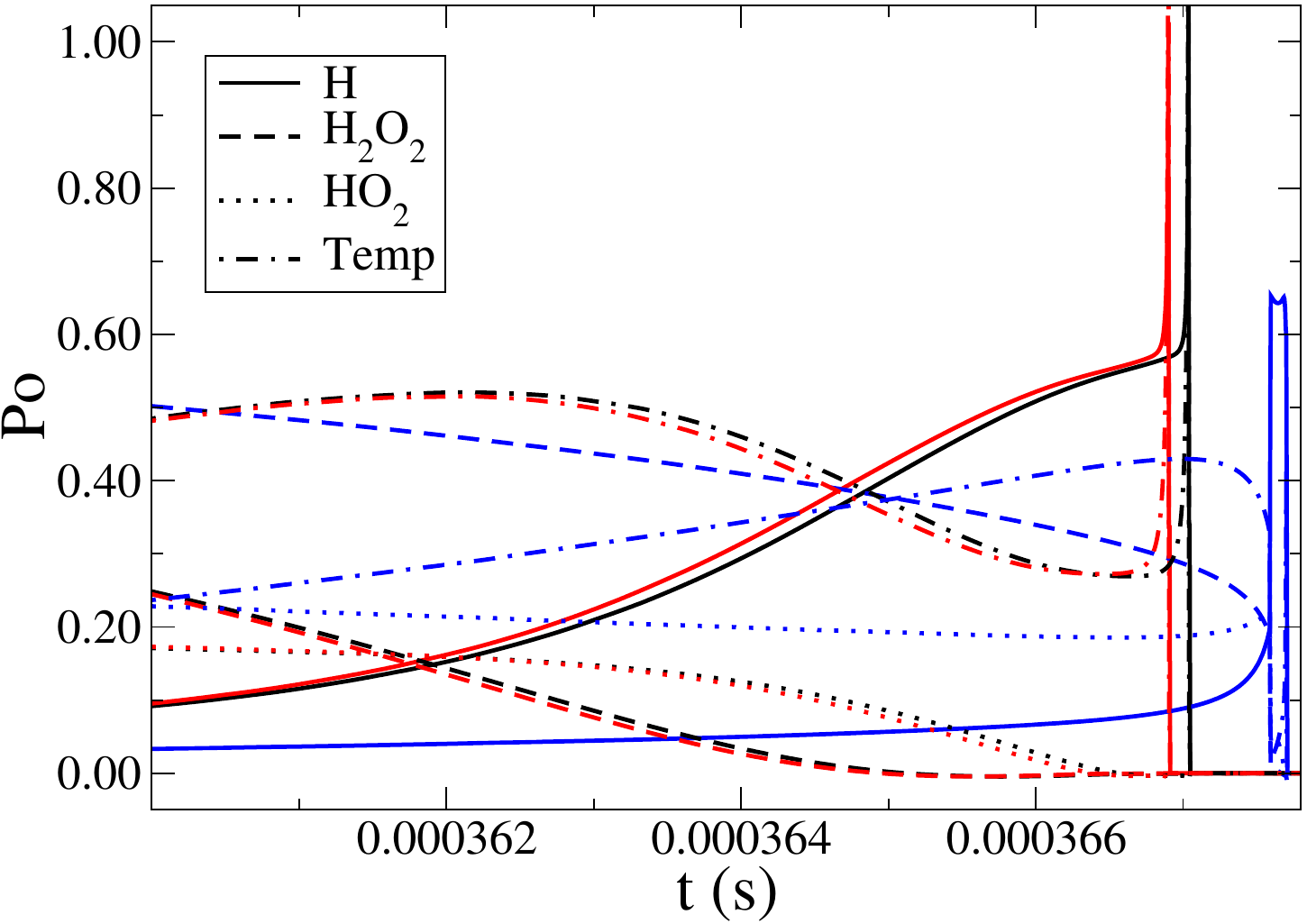}
\caption{}
\label{fig:Po3}
\end{subfigure}
\hfill
\begin{subfigure}[b]{0.45\textwidth} 
\centering
\includegraphics[scale=0.25]{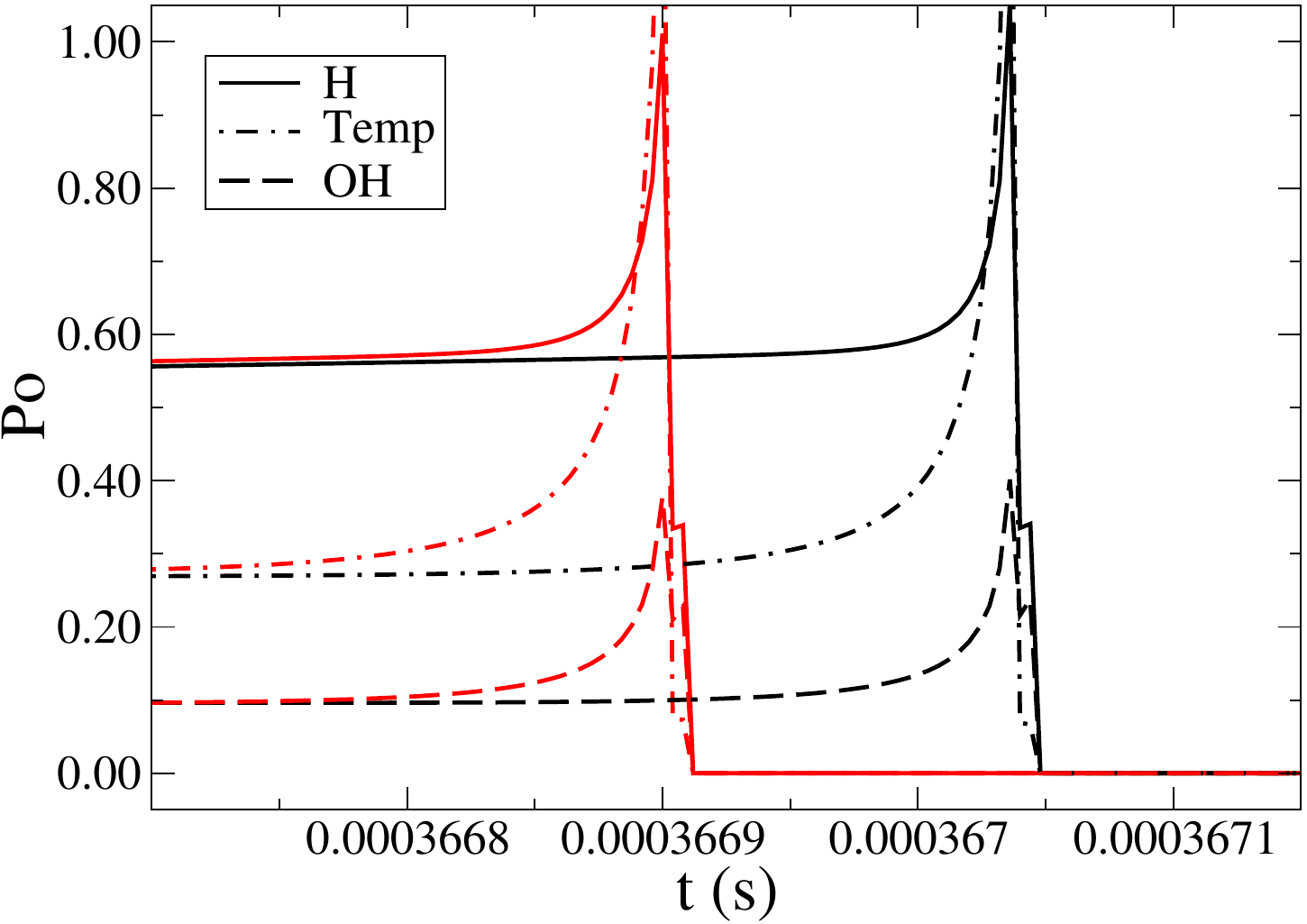}
\caption{}
\label{fig:Po4}
\end{subfigure}\hfill\hfill
\caption{The evolution of the variable that are most related to the explosive mode, for the 3 points under investigation. Black: x$_1$, red: x$_2$ and blue: x$_3$.}
\label{fig:Po}
\end{figure}

First stage: from 0 $\mu$s to $\sim$240 $\mu$s, where the values of the relative contribution of each pointed variable is maintained, as seen in Fig. \ref{fig:Po1}. Second stage: form $\sim$240 $\mu$s to $\sim$366 $\mu$s, where important changes in the pointed variables are taking place, as well as the transition to thermal runaway regime, as it is manifested in Fig. \ref{fig:Po3}. Finally, the 3rd stage, which characterizes the end of the explosive stage, after 366 $\mu$s, where the autoignition is completed (Figs.~\ref{fig:Po3} and ~\ref{fig:Po4}).

Considering the pointed variables of each stage, for the first stage and for all cases, only H$_2$O$_2$ and HO$_2$ species are found to be related the most to the fast explosive mode.~For all cases, HO$_2$ contributes around 70$\%$ throughout the first stage, while the rest 30$\%$ comes from H$_2$O$_2$.~Both species are found to be reactants at two of the reactions that are identified by TPI to contribute the most to the explosive time scale at this stage in all cases; i.e. reactions 20b and 18f.~For the second stage, the relative contribution of the 1$^{st}$ pointed HO$_2$ starts to decrease, slowly at first and faster afterwards, accompanied by an increase of the relative contribution of H$_2$O$_2$ up to 60$\%$ which then follows a steeper than the HO$_2$ decrease.~This behavior is found to be faster in case 1, as a result to the fact the x$_1$ is in a more reactive state than the others, at this point.

In the mean time, temperature contribution has started to risen, taking the lead, which around $\sim$362 $\mu$s it reaches more than 50$\%$ and becomes the most significant contributor to the explosive dynamics, declaring the transition to thermal runaway regime (Fig.~\ref{fig:Po3}).~Furthermore, the temperature rise, at this time for case 2, has already become larger than the other cases (Fig.~\ref{fig:mean}).~Due to this, the highly exothermic reactions 1f and 13f are benefited, as declared in Fig.~\ref{fig:TPI3}, which both consume H to produce OH radicals.~This behavior is captured by the CSP Pointer at this stage, as shown in Fig.~\ref{fig:Po3}:~Being a reactant in the reactions of largest relative contribution to the explosive timescale, the relative contribution of H-radical starts to become important.~Even though the profiles of H$_2$O$_2$, HO$_2$ and temperature of case 2 follow the same behavior with case 1, but with smaller contributions, when starts to become significant, the relative contribution of H appears to be larger than the one of case 1, as shown in Fig.~\ref{fig:Po3}.~This is manifested right after temperature relative contribution reaches its local pick.~Thus, even though temperature contribution is less at case 2 compared to case 1, it is enough to increase the OH radical pool (Fig.~\ref{fig:species}) to the point that the transition to the end of the explosive stage and towards its autoigntion, is faster at case 2 than the other cases.

Finally, the increase of H relative contribution accompanying the decrease of the one of temperature continues until $\sim$366.9 $\mu$s, where a very  steep increase is manifested for both H and temperature.~This point in time is matching the time where $\tau_{e,f}$ accelerates until it reaches its minimum value, shown in Fig.~\ref{fig:exp3}, indicating the extremely fast dynamics.~At that point, both exhibit maximum values of contribution, along with OH contribution that appears to be significant, as well, right before the end of the explosive stage, towards autoignition (Fig.~\ref{fig:Po4}).~Note that the pointed OH is a reactant of reaction 4f that was previously identified by TPI as a reaction that promotes the explosive action of $\tau_{e,f}$.

\subsubsection*{CSP API}

\begin{figure}[h!]
\begin{subfigure}[b]{0.45\textwidth} 
\centering
\includegraphics[scale=0.25]{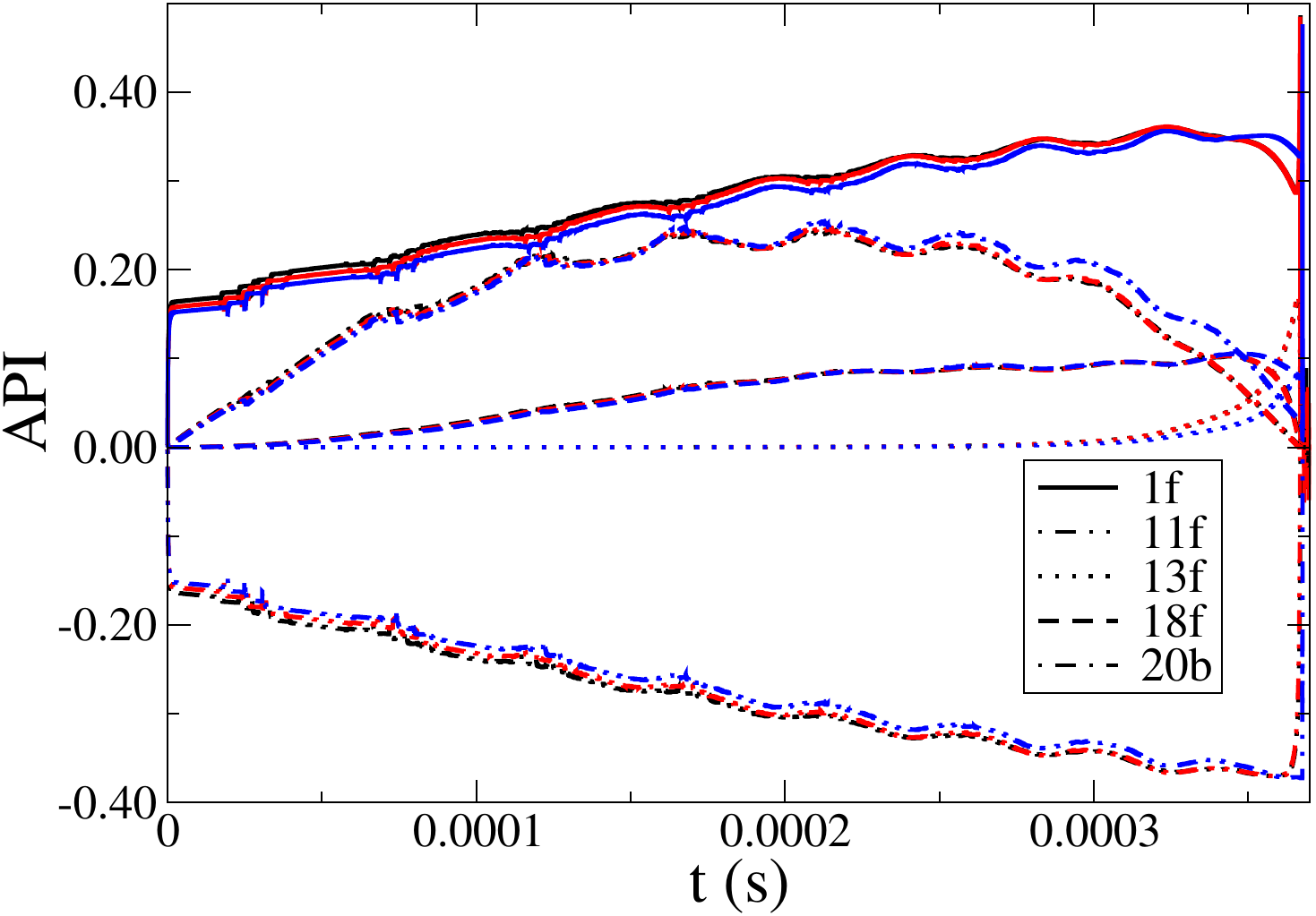}
\caption{}
\label{fig:API1}
\end{subfigure}
\hfill
\begin{subfigure}[b]{0.45\textwidth} 
\centering
\includegraphics[scale=0.25]{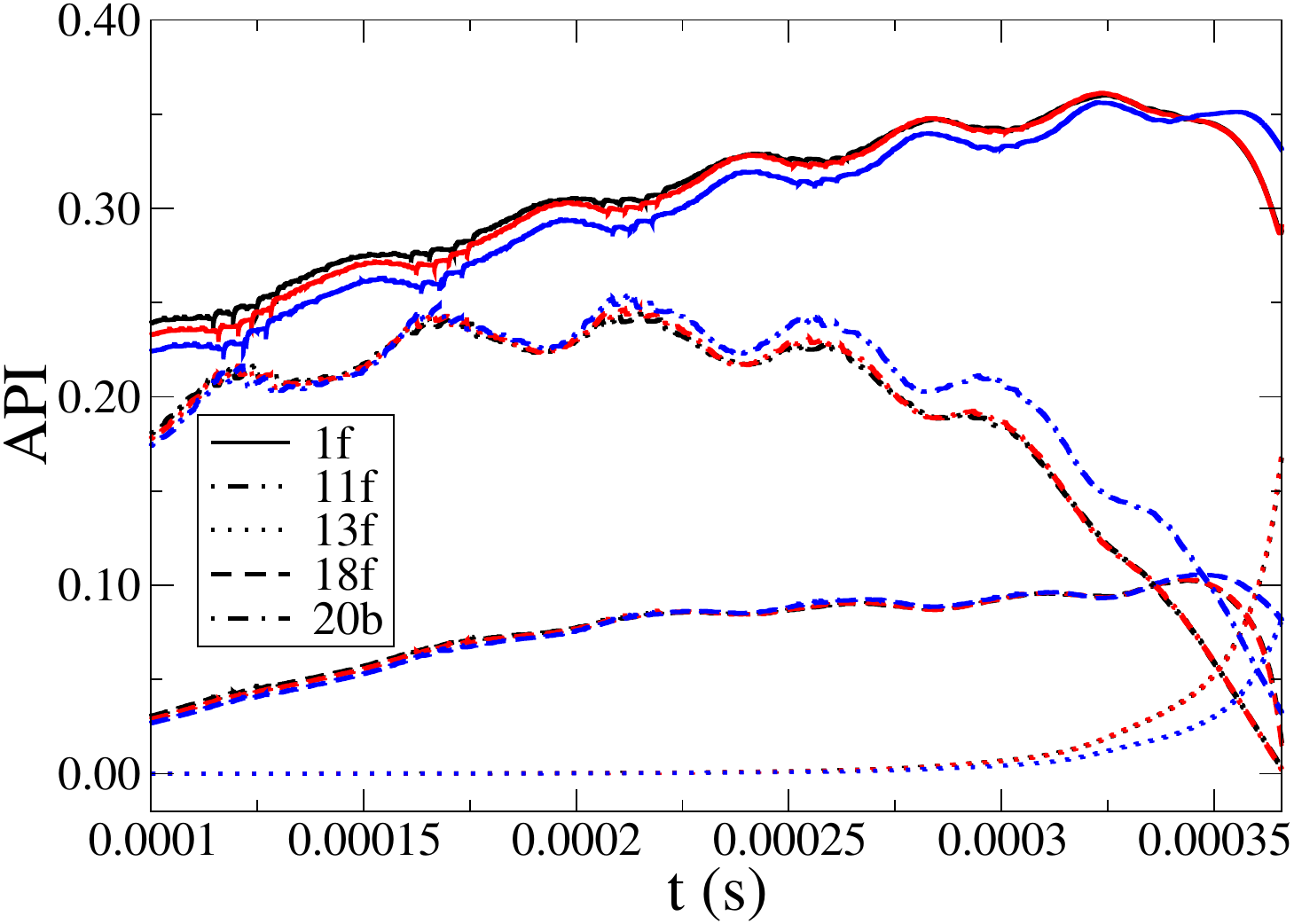}
\caption{}
\label{fig:API2}
\end{subfigure}\hfill\hfill
\\
\hfill\hfill
\begin{subfigure}[b]{0.45\textwidth} 
\centering
\includegraphics[scale=0.25]{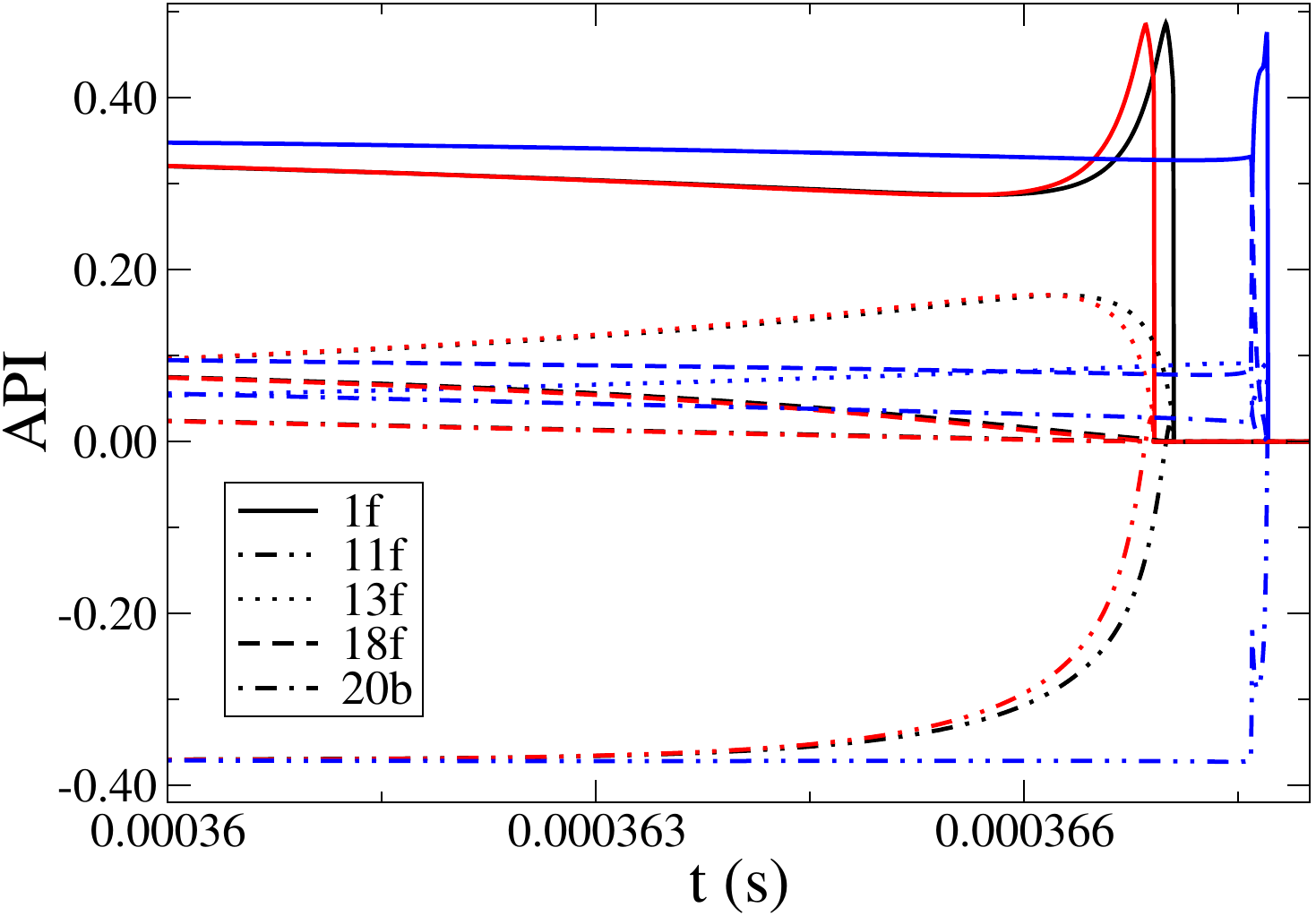}
\caption{}
\label{fig:API3}
\end{subfigure}
\hfill
\begin{subfigure}[b]{0.45\textwidth} 
\centering
\includegraphics[scale=0.25]{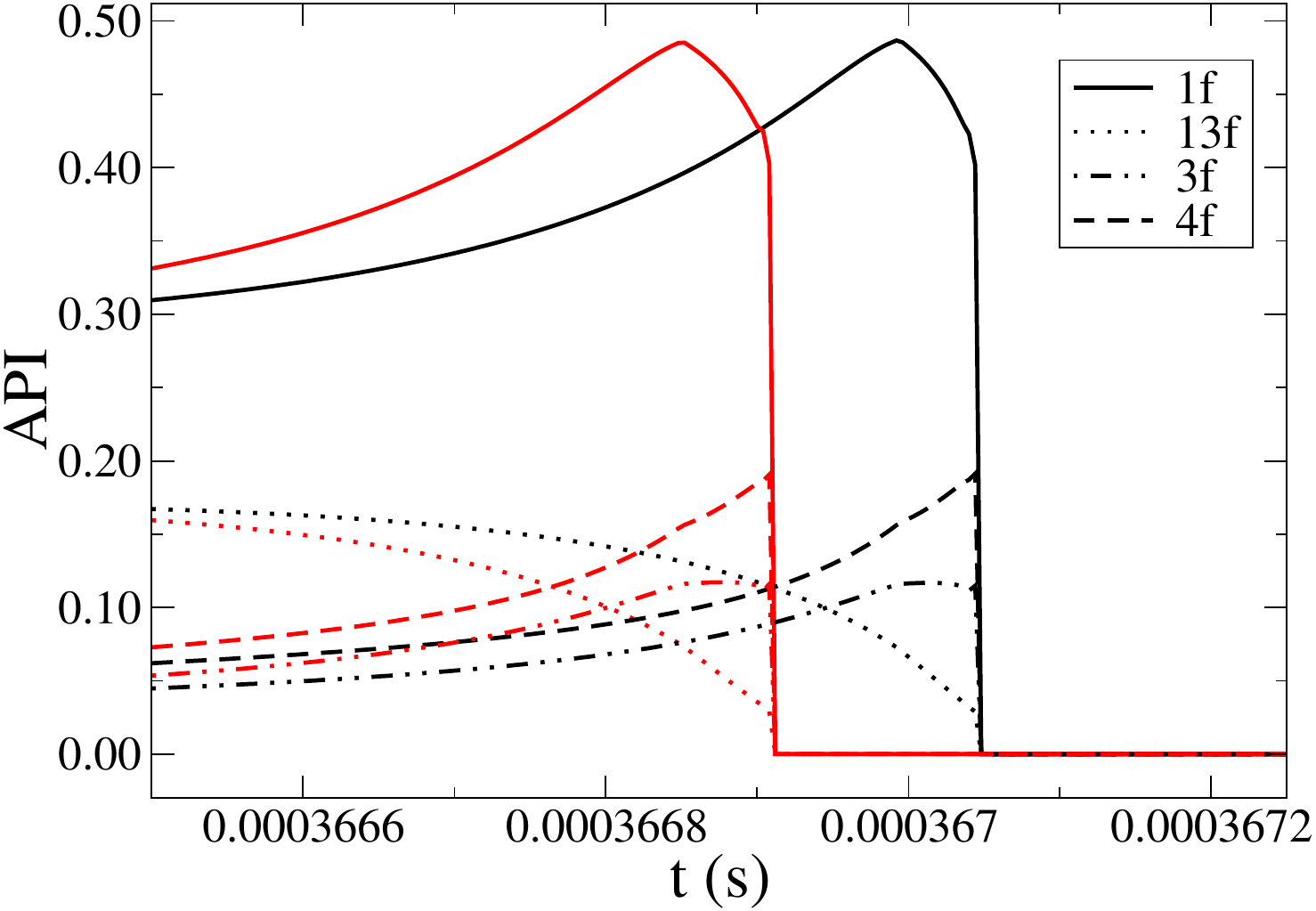}
\caption{}
\label{fig:API4}
\end{subfigure}\hfill\hfill
\caption{The evolution of the largest values of API, for the 3 tpoints under investigation. Black: x$_1$, red: x$_2$ and blue: x$_3$.}
\label{fig:API}
\end{figure}

In order to study the effect of each process to the fast explosive mode, by identifying the reactions that tend to increase or decrease its impact to the evolution of the system, the \textit{CSP Amplitude Participation Index} was used.~In this study, only the chemical source term from the RHS of Eq.~\eqref{eq:gov1a} was taken under consideration for the calculation of the CSP amplitudes, so only the effect of reactions was studied.~The reactions that contribute the most to the amplitude of the fast explosive mode, f$^{e,f}$, were identified throughout the solution and their relative contribution is displayed in Fig.~\ref{fig:API}.

%

For all cases, the same set of reactions that contribute to the fast explosive timescale, were also identified to contribute to the amplitude f$^{e,f}$, for most of the process, as well, as shown in Fig.\ref{fig:API1}.~More particular, reaction 1f is the major contributor that tends to increase f$^{e,f}$.~While the process evolves, its relative contribution increases, similarly to the TPI profile that was described in a previous section.~However, the main opposition to f$^{e,f}$ comes from reaction 11f, which competes with the action of reaction 1f throughout the process.~This results in canceling the action of reaction 1f up to $\sim$365 $\mu$s, where the relative contribution of reaction 11f starts to drop (Fig.\ref{fig:API3}), in contrast to reaction 1f which exhibits a steep increase towards the end of the process.

For this first stage of 1f-11f cancellation, the major relative contribution comes from the second largest contributor, reaction 20b.~Its relative contribution starts from zero at the beginning of the process and is increased until the middle of this stage, around $\sim$190 $\mu$s, where the slow explosive time scale appears.~Then its relative contribution starts decreasing until it drops to zero at the end of the explosive stage.~Note that the profiles of both 1f and 20b reactions are representative of the effect of the pressure, as shown in Fig.~\ref{fig:API2} and more specifically, the contribution of reaction 20b appears to be very sensitive to the pressure wave interaction.

For the rest of the minor contributions, the profiles of Fig.~\ref{fig:API} indicate that each reaction follows the same contribution within the time frame of their action as indicated by the TPI, for all cases.~Reaction 13f, which is the second largest contributor when the contribution of reaction 20b has been exhausted, along with reaction 18f, tend to increase more the f$^{e,f}$ (Fig.~\ref{fig:API3}), for all cases.~Close to ignition, when the relative contribution of reaction 20b drops lower that 0.1, as Fig.~\ref{fig:API4} indicates, the contribution of reactions 3f and 4f become the important complementary to reaction 1f contributors to f$^{e,f}$ increase.




\subsubsection*{Effect on the rate constants}
\label{Average}


\begin{figure}[h!]
\hfill\hfill
\hspace{-5mm}
\begin{subfigure}[b]{0.31\textwidth} 
\includegraphics[scale=0.4]{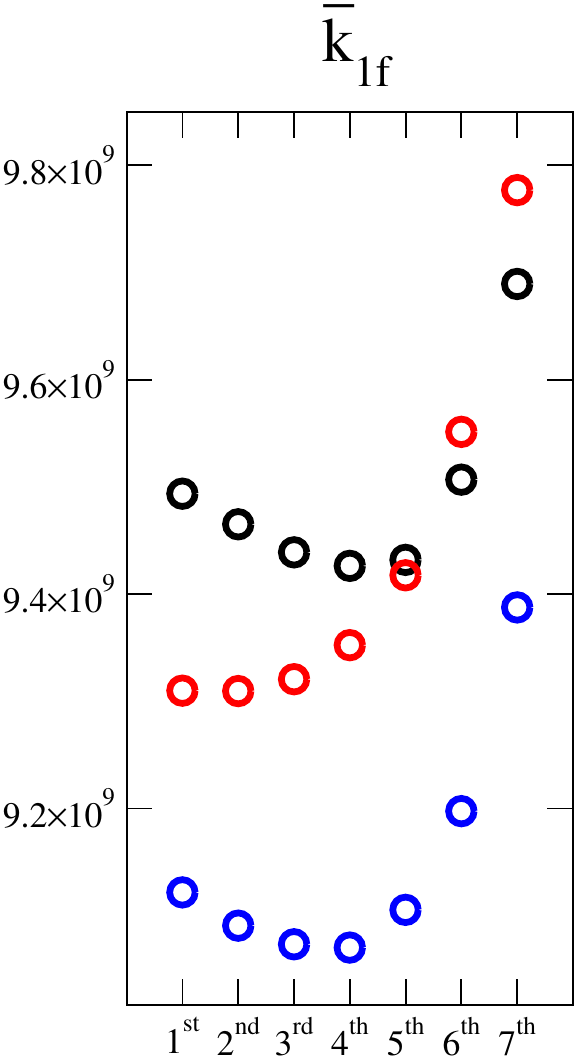} 
\caption{}
\label{fig:rates1}
\end{subfigure}
\hfill
\begin{subfigure}[b]{0.31\textwidth} 
\includegraphics[scale=0.4]{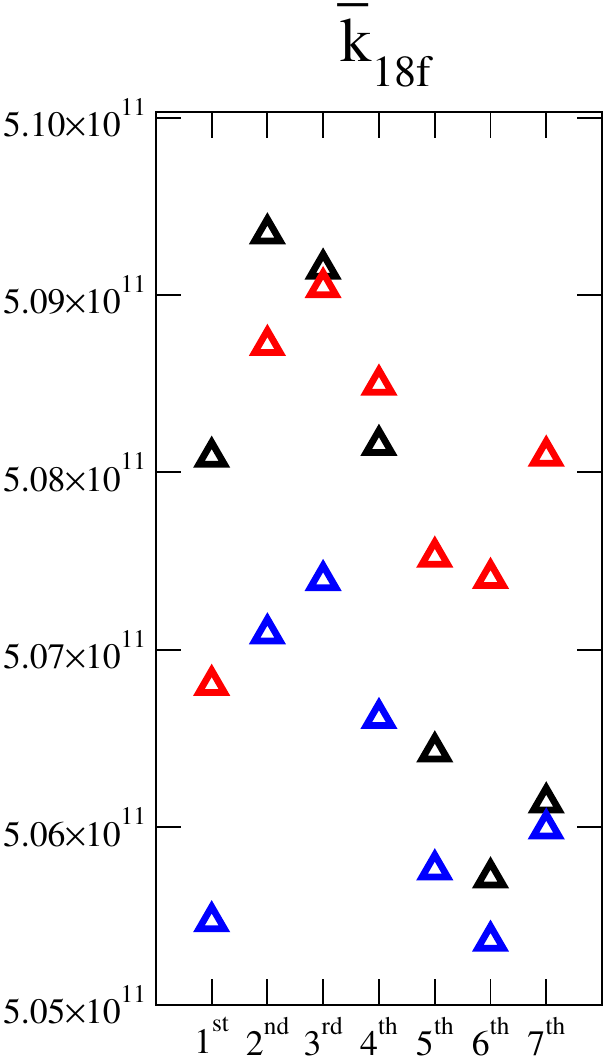}
\caption{}
\label{fig:rates2}
\end{subfigure}
\hfill
\begin{subfigure}[b]{0.31\textwidth} 
\includegraphics[scale=0.4]{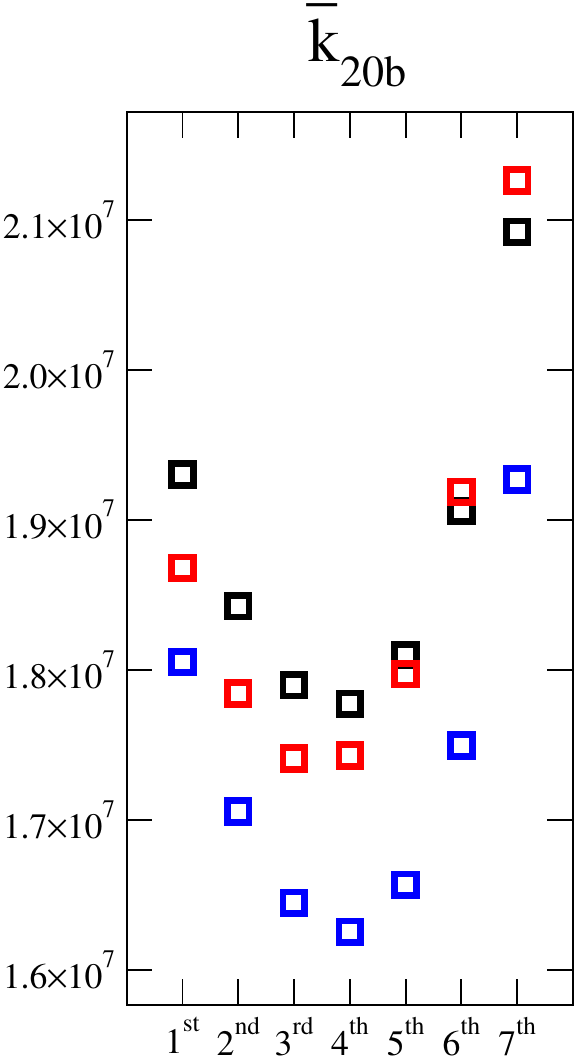} 
\caption{}
\label{fig:rates3}
\end{subfigure}
\hfill
\caption{The average rate constants of the reactions that promote the most the explosive character of $\tau_{e,f}$, for each period, for the 3 points under investigation. Black: x$_1$, red: x$_2$ and blue: x$_3$.~The values have been normalized in every period, similar to Fig.~\ref{fig:mean}, so that the relative differences to be highlighted.}
\label{fig:rates}
\end{figure}

Even though it is shown that the action of the important reactions, as identified by the CSP diagnostic tools, is responsible for the autoignition of x$_2$ at only the last part of the explosive stage, it is found that throughout the solution, the radical built up in case 2 was relatively accelerated compared to case 1 that lead to its autoignition.

The values of the rate constants of all reactions where calculated and it was found that the most significant changes appeared at the reactions identified by the CSP analysis to affect the most both the time frame of action and the impact of the explosive mode to the evolution of the system.~For these reactions, i.e. 1f, 18f and 20b, the rate constants were calculated for each period and are displayed in Fig.~\ref{fig:rates}.~In particular, the average values have been plotted for every period, normalized in a similar to Fig.~\ref{fig:mean} way, in order to highlight the relative differences between the three cases on each period.

The relative acceleration of the dynamics of case 2 is manifested in all rate constants of the most important reactions.~As expected, at the beginning of the process, the reaction rates of case 1 are larger than the ones of case 2, which, in turn, are larger than those of case 3.~In time, the net values of case 2 start increasing more than the relative increase of cases 1 and 3.~This is a result of the relative increase of the temperature of each case, as manifested in Fig.~\ref{fig:mean}.~The rate constant that is affected the most is the one of reaction 18f (Fig.\ref{fig:rates2}), where, from the 4th period already, case 2 has prevailed.~Reaction 18f is a third body reaction, which (third body) is indicative of the role of pressure to the system.~Since pressure effects are the driving mechanism of the detonation problem, it is expected that such a reaction to be affected the most.~However, despite its large values, its contribution to the explosive mode of the system is not the biggest, as TPI (Fig.~\ref{fig:TPI}) indicates.~Instead, the rate constant of reaction 1f (Fig.~\ref{fig:rates1}) is found to represent more smoothly the transition to case 2 dominance.~Finally, the rate constant of reaction 20b follows with smaller values and smaller contribution (Fig.~\ref{fig:rates1}).~In the last two reactions, case 2 is found to become dominant after the 5th period.



\section{Conclusions}
\label{Conclusions}
By using the CSP diagnostic tools, the case of detonation development from a pre-ignition front on a one-dimensional constant-volume stoichiometric hydrogen/air reactor with temperature stratification was analyzed.~A flame was initiated near the left wall and the effects of its propagation towards the cold end-wall with negative temperature gradient were analyzed.~It was found that ignition occurs first in the location within the cold-spot, ahead of the flame and far from the end-wall, which then evolves to a detonation.

A set of CSP diagnostics were obtained in three indicative points in space, within the cold spot region, in the vicinity of the point where the autoignition occurs:~The first point, x$_1$ was selected in the region right after the start of the temperature stratification which exhibited the most reactive state, the second point, x$_2$, is closer to the end-wall and the one where autoignition occurs and the third point, x$_3$, was selected so that x$_2$ to be equidistant from x$_1$ and x$_3$, to the direction of the end-wall, were autoignition occurs in the cases of positive temperature gradient.~The mechanism by which the generated pressure waves influence the autoignition within the cold-spot was investigated.~Furthermore, the reactions that support or oppose the generation of the explosive dynamics and the species that relate the most to the explosive modes at the three points were identified and the relation with the oscillating pressure build-up was studied. 

It was found that the pressure oscillations tend to synchronize in the chamber, resulting in a local pressure build-up in a periodic manner. The average pressure at each period was found to be similar for all three cases, while the average temperature values at each period is greater at x$_2$, compared to the rest points. As a result, the rate constants of the most important reactions were found to become larger there, leading to a more reactive state that accelerates the dynamics of x$_2$ and to its autoignition.

The CSP diagnostic tools identified three reactions to contribute the most to the generation of the explosive modes, both by supporting its explosive character and by increasing its impact to the evolution of the system, for all three cases: 1f: H + O$_2\rightarrow$ O + OH, 18f: H$_2$O$_2$ (+M) $\rightarrow$ OH + OH (+M) and 20b: H$_2$O$_2$ + H $\leftarrow$ HO2 + H2.~All these reactions contribute to the production and support of the H and OH radical pools needed for the ignition process.~The specific role of H and OH radicals throughout the process and the action of each one was identified and analyzed.~Compared to case 1, which exhibits initial and for most of the process evolution higher reactivity, it was found that, even though the thermal runaway regime lasts the for same amount of time in both cases, its effect to the explosive mode and to the variables related the most to it, is stronger in case 2.~In particular, for case 2, it was found that OH is affected more by the temperature oscillations, resulting to the increase of its radical pool, which leads to a faster ignition.~The increased concentration of H and OH radicals and their strong relation to the explosive modes in case 2 lead to the conclusion that, even though x$_2$ initially lies in a position of a less reactive state, the pressure-temperature built up results in an acceleration of its dynamics.

Considering case 3, it was clear from all CSP diagnostics that x$_3$ is affected only by the pressure wave of the autoignitive x$_2$ and subjected to the detonation initiation.~Even though ignition manifests gradually along cases 1 and 2, in case 3 it manifests abruptly and no radical built up is indicated.~Instead, ignition of case 3 manifests sooner than expected and in a much more intense manner, as shown in Fig.\ref{fig:exp3}.~As it was stated earlier, this feature is due to the fact that ignition at case 3 is caused by the action of the advancing pressure front and not by the action of heat releasing chemical reactions.

\subsection*{Future Work}
Since this analysis was concentrated on the effect of the chemistry involved on the evolution of the system, the next step is to study the effect of transport to the dynamics.~By accounting also for all the diffusion and convection terms of the system, the CSP analysis provides an algorithmic methodology that can determine the role of the individual processes to the evolution of the system, whether these are diffusive/convective or chemical kinetics processes.~Additionally, by identifying the dominant physical processes, we can extract the key sub-processes underlying the important explosive dynamics that are responsible for the transition to detonation and obtain this way a better physical understanding of the problem.

Furthermore, the possibility of using additives as a way of manipulating the chemical pathways through which the system evolves can be studied.~This way, through the systematic approach of targeted species' addition, the avoidance or control of knock occurrence can be obtained.

\section*{Acknowledgments}
\label{Acknowledgments}
This work was sponsored by competitive research funding from King Abdullah University of Science and Technology (KAUST). The support from Khalifa University of Science and Technology, via project 8434000269, is gratefully acknowledged.


\bibliography{Bibliography.bib} 
\bibliographystyle{unsrtnat}






\end{document}